\documentclass[usegraphicx,useAMS,usenatbib]{mn2e}

\newcommand{\lii}{Li\,{\footnotesize I}}

\newcommand{\kms}{\,km\,s$^{-1}$}

\newcommand{\be}{\begin{equation}}
\newcommand{\ee}{\end{equation}}
\newcommand{\bd}{\begin{displaymath}}
\newcommand{\ed}{\end{displaymath}}

\title[Low mass PMS stars in NGC 2169]
  {The Keele-Exeter young cluster survey: I. Low mass pre-main sequence stars in NGC 2169}
\author[R.D. Jeffries et al.]
  {R.D.~Jeffries$^1$, J.M.~Oliveira$^1$, Tim Naylor$^2$, N.J. Mayne$^2$ and
  S. P. Littlefair$^{2,3}$\\
  $^1$ Astrophysics Group, School of Physical and Geographical Sciences, Keele University, Keele, 
      Staffordshire ST5 5BG\\
  $^2$ School of Physics, University of Exeter, Stocker Road, Exeter
  EX4 4QL\\
  $^3$ Department of Physics and Astronomy, University of Sheffield, Sheffield S3 7RH\\
}
\date{Submitted 5 September 2006}

\pagerange{\pageref{firstpage}--\pageref{lastpage}} \pubyear{2004}

\def\LaTeX{L\kern-.36em\raise.3ex\hbox{a}\kern-.15em
    T\kern-.1667em\lower.7ex\hbox{E}\kern-.125emX}

\begin{document}

\label{firstpage}

\maketitle

\begin{abstract}
We have used $R_C I_C$ CCD photometry from the Isaac Newton telescope
and intermediate resolution spectroscopy from the Gemini North
telescope to identify and characterise low-mass
($0.15<M/M_{\odot}<1.3$) pre-main sequence stars in the young open
cluster NGC 2169. Isochrone fitting to the high- and low-mass
populations yields an intrinsic distance modulus of
$10.13^{+0.06}_{-0.09}$\, mag and a model-dependent age of $9\pm
2$\,Myr.  Compared with the nearby, kinematically defined
groups of a similar age, NGC~2169 has a large
low-mass population which potentially offers a more precise statistical
investigation of several aspects of star formation and early stellar
evolution.  By modelling the distribution of low-mass stars in the
$I_C$ versus $R_C-I_C$ diagram we find that any age spread among
cluster members has a Gaussian full width at half maximum $\leq
2.5$\,Myr. A young age and small age spread ($<10$\,Myr) are supported
by the lack of significant lithium depletion in the vast majority of
cluster members. There is no clear evidence for accretion or warm
circumstellar dust in the low-mass members of NGC 2169, bolstering the
idea that strong accretion has ceased and inner discs have dispersed in
almost all low-mass stars by ages of 10\,Myr.
\end{abstract}


\section{Introduction}

There are notably few well-studied clusters in the literature with ages
between about 5 and 30\,Myr.  Yet investigating the coeval populations
in such clusters is vital for our understanding of: (i) the lifetimes
and subsequent evolution of high mass (8--20\,$M_{\odot}$) main sequence
stars which contribute the majority of metal enrichment to the
universe; (ii) the evolution of circumstellar material, formation of
planetary systems and loss of angular momentum in lower mass stars.

In this paper we report the first results from the Keele-Exeter young
cluster survey (KEY clusters) to find and
characterise clusters with age 5--30\,Myr. NGC~2169 ($=$C\,0605+139) is
a concentrated, but sparsely populated young open cluster (class I3p --
Ruprecht 1966) in the constellation of Orion, with an age of $\simeq
10$\,Myr and distance of $\sim 1$\,kpc (see Perry, Lee \& Barnes 1978 and
Section~\ref{previous}).  Using CCD photometry and intermediate
resolution spectroscopy we have uncovered the low-mass
($0.1<M/M_{\odot}<1.3$) pre-main sequence (PMS) 
population of NGC~2169 and used isochronal fits to the
high- and low-mass stellar populations and measurements of photospheric
Li depletion to test evolutionary models, determine the cluster age and
investigate the possibility of any age spread within the cluster. Armed
with an age, we have investigated timescales for the dispersal of gas and
dust discs by comparing the low mass stars of NGC~2169 with those in
younger and older clusters.

The paper is organised as follows.
Section~2 summarises previous work on this cluster; Section
3 describes a new photometric survey (the results of which are provided
in electronic form) used to identify candidate
low-mass cluster members; Section 4 describes intermediate resolution
spectroscopy from the Gemini North telescope which is used to confirm
candidate membership and study Li depletion; Section 5 compares
constraints on the cluster age imposed by low-mass isochrones, Li
depletion models and the evolutionary status of high-mass stars in the
cluster; Section 6 deals with the
spatial structure and mass function of the newly discovered low-mass
population; Section 7 discusses the evidence for age spreads within the
cluster and the evolution of circumstellar accretion. 
Our conclusions are presented in Section 8.

\begin{table*}
\caption{Literature evaluations of the age, distance and reddening of
  NGC 2169}
\begin{tabular}{lcccl}
\hline
Authors       &  $E(B-V)$  & Intrinsic Distance  &  Age  & Notes \\
              & (mag)      & Modulus (mag)       & (Myr) &       \\
\hline
Sagar (1976)  & 0.18       &  9.60               & $<9$     & $UBV$
photoelectric\\
Harris (1976) &            &                     & $<12$   & MK spectra \\
Abt (1977)    & $0.17\pm0.03$ & $10.9\pm0.3$     &         & MK spectra \\
Perry et al. (1978) & $0.20\pm0.01$ & $10.2\pm0.1$ & $<23$ &
$ubvy\beta$ photoelectric\\
Delgado et al. (1992) & 0.20 &  10.05            & $<16$   &
$ubvy\beta$ photoelectric\\ 
Pe\~na \& Peniche (1994) & $0.25\pm0.02$ & $9.7\pm 0.3$ & $<50$ &
$ubvy\beta$ photoelectric \\
\hline
\end{tabular}
\label{clusterparams}
\end{table*}

\begin{figure*}
\includegraphics[width=150mm]{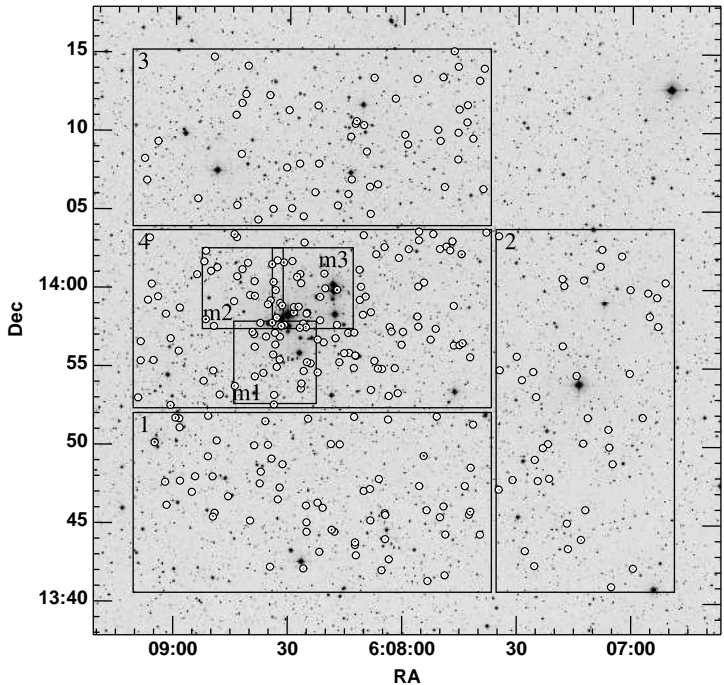}
\caption{Location of the INT CCD and GMOS spectroscopic surveys. The
  large rectangles (labelled 1 to 4) show the individual detectors from
  the INT WFC (see Section~\ref{ccdphotom}). 
  The small squares (labelled m1 to m3) show the three
  GMOS multislit masks (see Section~\ref{gmosobs}). The small symbols
  mark the locations of photometric cluster candidates with
  $15.5<I_C<19.0$ (see Section~\ref{clustermass}).
}
\label{gaiaplot}
\end{figure*}

\section{Previous studies of NGC 2169}

\label{previous}

There are several previous studies of the high mass
population in NGC~2169.
Photoelectric photometry of the brighter candidate cluster
members is presented in Cuffey \& McCuskey (1956, in the $PV$ system),
Hoag et al. (1961, $UBV$), Sagar (1976, $UBV$), Perry et al.
(1978, $uvby\beta$), Delgado et al. (1992,
$uvby\beta$) and Pe\~na \& Peniche (1994, $uvby\beta$). Harris
(1976) and Abt (1977) have published spectral types on the
Morgan-Keenan system for the brightest
stars in the cluster and more recently, Liu, Janes \& Bania (1989)
conducted a radial velocity (RV) survey of 9 bright A and B stars in
the cluster, finding several spectroscopic binaries. A number of
surveys for peculiar or variable stars have been conducted (see
Jerzykiewicz et al. 2003 and references therein) and at least two
candidate beta Cepheids and an A0V Si star have been identified.

For this paper, the important parameters are the cluster age, distance
and reddening. Table~\ref{clusterparams} gives a summary of the
conclusions reached by other authors. There is good agreement between
photometric and spectroscopic determinations of the reddening and very
little star-to-star dispersion ($\sigma_{E(B-V)}\leq 0.02$ -- Delgado
et al. 1992). There is less agreement in the cluster distance and
age, primarily because of disagreement over which stars should be
considered members of the cluster and which stars are binary systems,
but also because of different calibrations of the main sequence
turn-off upon which the age estimates are based. As there are no
obviously evolved stars apart from the binary system Hoag 1 (B2III --
Abt 1977) these ages are inevitably upper limits.  We prefer the
estimates provided by Perry et al. (1978) and Delgado et al. (1992),
which attempt to weed out field interlopers before estimating the
cluster parameters. For the moment we assume the cluster is
less than 23\,Myr old, at a distance of $\simeq 1000$\,pc and has $E(B-V)=0.20$, although
these parameters are re-assessed in Section~\ref{age}.

\begin{figure}
\includegraphics[width=75mm]{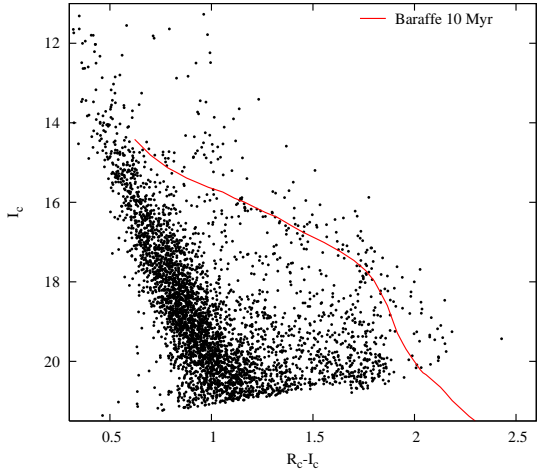}
\caption{A colour-magnitude diagram 
for unflagged objects with uncertainties $<0.1$ mag in $I_C$ and
$R_C-I_C$ seen in CCD 4 (see Fig.~\ref{gaiaplot}).
The solid line shows a theoretical 10\,Myr PMS (from Baraffe et
al. 2002) at an intrinsic distance modulus
of 10.13, and with a reddening/extinction corresponding to $E(R_C-I_C)=0.14$
(see Section~\ref{age}).
}
\label{cmd1}
\end{figure}

\begin{table}
\caption{
The range of colours for the photometric standards observed by each CCD detector.
}
\begin{tabular}{ccccccccccccccccc}
\hline
CCD     & Colour range \cr 
\hline
1       &  0.340 $< R_C-I_C < $ 1.839  \cr
2       &  0.318 $< R_C-I_C < $ 1.750  \cr
3       &  0.342 $< R_C-I_C < $ 2.323  \cr
4       &  0.207 $< R_C-I_C < $ 2.314  \cr
\hline
\end{tabular}
\label{stand_col}
\end{table}

\section{CCD photometry of NGC 2169}

\label{ccdphotom}

We observed NGC~2169 on the night beginning 28th September 2004, using the
Wide-Field Camera (WFC) on the 2.5-m Isaac Newton telescope (INT).
The camera consists of 4 thinned EEV 2kx4k CCDs (numbered 1--4) covering 0.33
arcsec/pixel on the sky. The arrangement of the 4 detectors on the sky
for our observations of NGC~2169 is shown in Fig.~\ref{gaiaplot}.
We obtained exposures in the Cousins $R$-band (3s, 30s and 3$\times$350s) 
and Sloan $i$-band (2s, 20s and 200s).
The night was photometric, and so we also obtained observations of
standard stars from 
Landolt (1992) and Stetson (2000).
Table \ref{stand_col} shows the range in colour of standards observed
for each CCD.

The data were de-biassed and flatfielded using master bias and master
twilight sky flat frames.
The $i$-band data were defringed using a library fringe frame.
We extracted the NGC~2169 photometry
using the optimal photometry techniques described in
Naylor (1998) and Naylor et al. (2002). 
We first searched the sum of all three long $i$-band frames to produce a
catalogue of objects, and then performed optimal photometry at the 
positions of these stars in all the frames.
We flagged as ``I'' (ill-determined sky) any stars which gave 
$\chi^2>3$ when we fitted the distribution of counts in the sky
(see Burningham et al. 2003).
By comparing the measurements in the long $R$ frames we established that we 
should add a one percent magnitude-independent uncertainty to measurements
from a single frame.
We included this when we combined the measurements to yield a single magnitude
for each star in each filter, and at the same time flagged as variable
any stars which had a reduced $\chi^2$ of 10 or more compared with a
constant value. The optimal photometry magnitudes were corrected to
that of a large aperture using a spatially dependent aperture
correction (see Naylor et al. 2002).

Standard star photometry was also extracted using optimal photometry
techniques and corrected to a larger aperture in the same way as the
target data. The advantage of this over the more usual method of
performing photometry directly in a large aperture was that we
collected good signal-to-noise photometry on many more faint standards.
The only disadvantage might be that some standards have been originally
defined using a large aperture that included other objects. However we
note that our reduction process flags objects with photometry that is
significantly perturbed by nearby companions and also that many of the
fainter standards (from Stetson 2000) were in fact measured using
PSF-fitting in any case.

We fitted our standard star observations as a function of colour and
airmass.
The airmass range of our standard stars is small (1.1 to 1.3) and close to the
airmass of our target observations (1.1), and so we fixed the
extinction.
Although a single linear relationship was sufficient to represent the
conversion from instrumental $i$ to
$I_C$ as a function of $R_C-I_C$, we found we had to use two separate
linear relationships to convert instrumental $R-i$ to $R_C-I_C$, with
the break occuring at $R_C-I_C=1.0$ to 1.3 depending on CCD.
We found that we needed to add a magnitude-independent uncertainty of 1 percent in
$R_C-I_C$ and 2 percent in $I_C$ (1.5 per cent if we just used the Landolt
standards) to obtain a reduced $\chi^2$ of unity.
These values therefore correspond to the combined uncertainty in our profile 
correction, and our correction to the Cousins system.
They are not included in the uncertainty estimates in our final catalogues, 
as they should not be added when comparing stars in a similar region of the
CCD (see Naylor et al. 2002).
We derived our astrometric calibration from 2MASS stars
(Cutri et al. 2003), with a RMS of 0.1
arcsec for the fit of pixel position as a function of RA and Dec.

Our entire catalogue is presented as Table \ref{ccd_catalogue}, which
is available on-line, or from the Centre de Donn\'ees astronomiques de
Strasbourg (CDS) or from the ``Cluster'' Collaboration's home
page\footnote{
http://www.astro.ex.ac.uk/people/timn/Catalogues/description.html}.  As
an example of the data, Fig.~\ref{cmd1} shows the $I_C$ vs $R_C-I_C$
colour-magnitude diagram (CMD) for all unflagged (i.e. clean, star-like
with good photometry) objects on CCD~4 with a signal-to-noise ratio
(SNR) greater than 10.  This CMD illustrates a clear PMS at the
position in the CMD appropriate for a $\sim 10$\,Myr population at a
distance of $\simeq 1000$\,pc. The sharp magnitude cut-off in
Fig~\ref{cmd1} is an artefact of the signal-to-noise threshold we have
placed on the plotted points. We judge our data to be almost complete
down to this cut-off, although the catalogue detection limit is about 1
magnitude fainter. For the purposes of this paper (see section~6) we
only require that the data is substantially complete to $I_C=19$ at
$R_C-I_C\simeq 1.9$.

\begin{table*}
\caption{
The $R_C$ vs $R_C-I_C$ photometric catalogue.
The full table is only available in electronic form, a portion is shown
here to illustrate its content. Columns list a unique identifier the
right ascension and declination (J2000.0), the CCD  pixel coordinates
at which the star was found, and then for each of $I_C$ and $R_C-I_C$
there is a magnitude, magnitude error and a flag (OO for a ``clean
detection -- a detailed description of the flags is given by Burningham
et al. [2003]).  
}
\begin{tabular}{cccccccccccc}
\hline
Cat & ID & RA & Dec & x & y& $I_C$ & $\delta I_C$ & flag & $R_C-I_C$ &
$\delta (R_C-I_C)$& flag \\
 &  &  \multicolumn{2}{c}{(J2000.0)} & & & \multicolumn{2}{c}{(mag)} & &
\multicolumn{2}{c}{(mag)} & \\
\hline
1.04&     206 & 06 08 32.011& +13 58 00.95&   1075.155&   1690.680&      9.043&      0.011&  SS&    1.016&      0.014&  SS\\
1.04&     225 & 06 08 30.184& +13 57 32.44&   1162.019&   1769.860&      9.222&      0.011&  SS&    0.922&      0.014&  SS\\
\hline
\end{tabular}
\label{ccd_catalogue}
\end{table*}

\section{Gemini Spectroscopy}

\subsection{Observations}

\label{gmosobs}

\begin{table*}
\caption{Gemini GMOS observation log giving the central position (telescope
  pointing) for each mask, the Julian Date at the observation midpoint, the
  number of targets in each mask, the exposure time and the average seeing.}
\begin{tabular}{ccccccc}
\hline
Mask & RA  & Dec & Date & N & Exp Time & Seeing\\
\#   & \multicolumn{2}{c}{J2000.0} & (JD-2453000) & & (s) & (arcsec) \\
\hline
1    & 06 08 34.8  & $+13$ 55 12   & 672.108& 18& $3\times 1800$& 0.6 \\
2    & 06 08 42.0  & $+14$ 00 00   & 706.973& 15& $3\times 1800$& 0.5 \\
3    & 06 08 24.4  & $+14$ 00 00   & 674.100& 15& $3\times 1800$& 0.5 \\  
\hline
\end{tabular}
\label{slitmask}
\end{table*}

\begin{figure*}
\centering
\begin{minipage}[t]{0.45\textwidth}
\includegraphics[width=71mm]{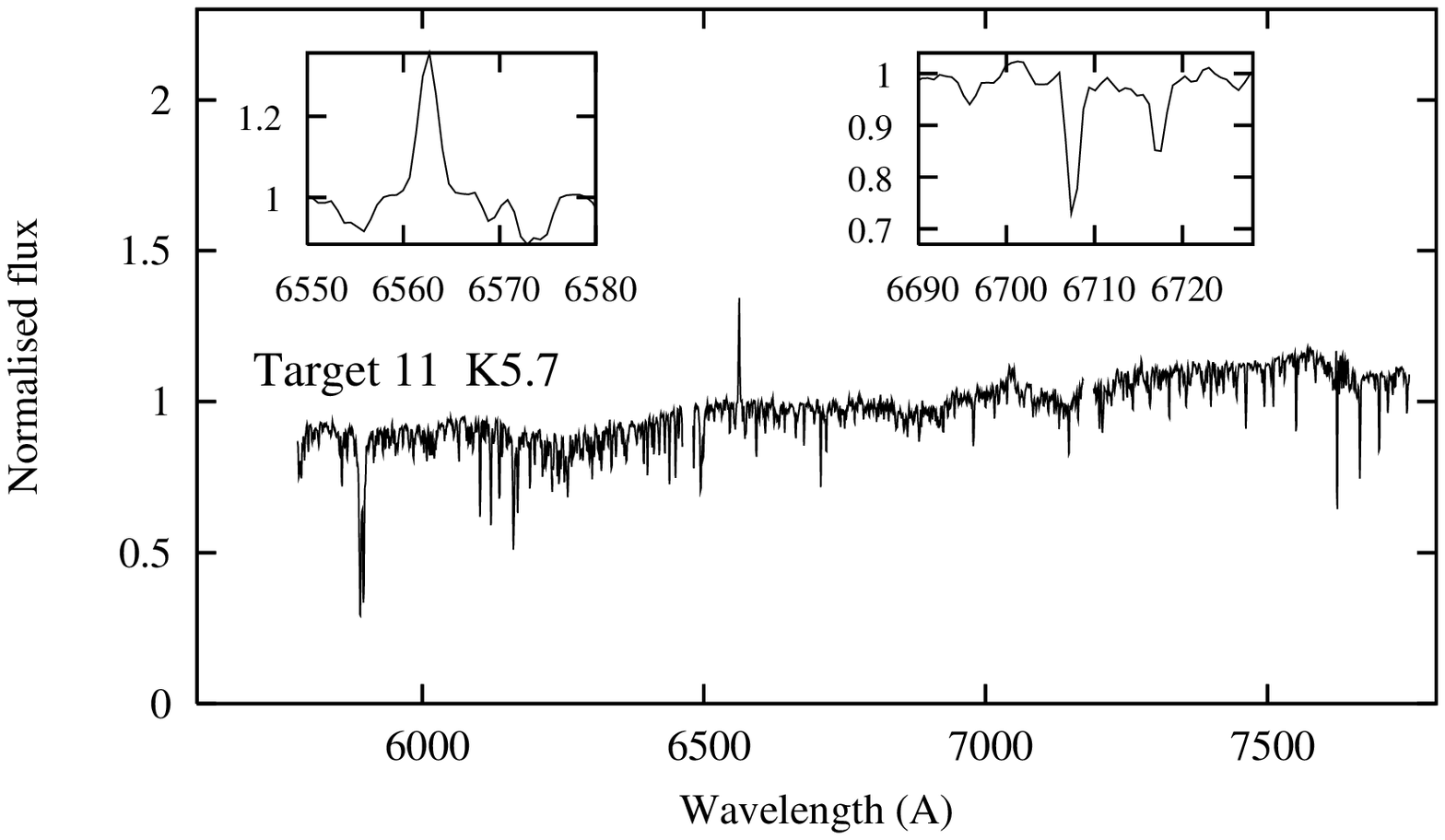}
\includegraphics[width=71mm]{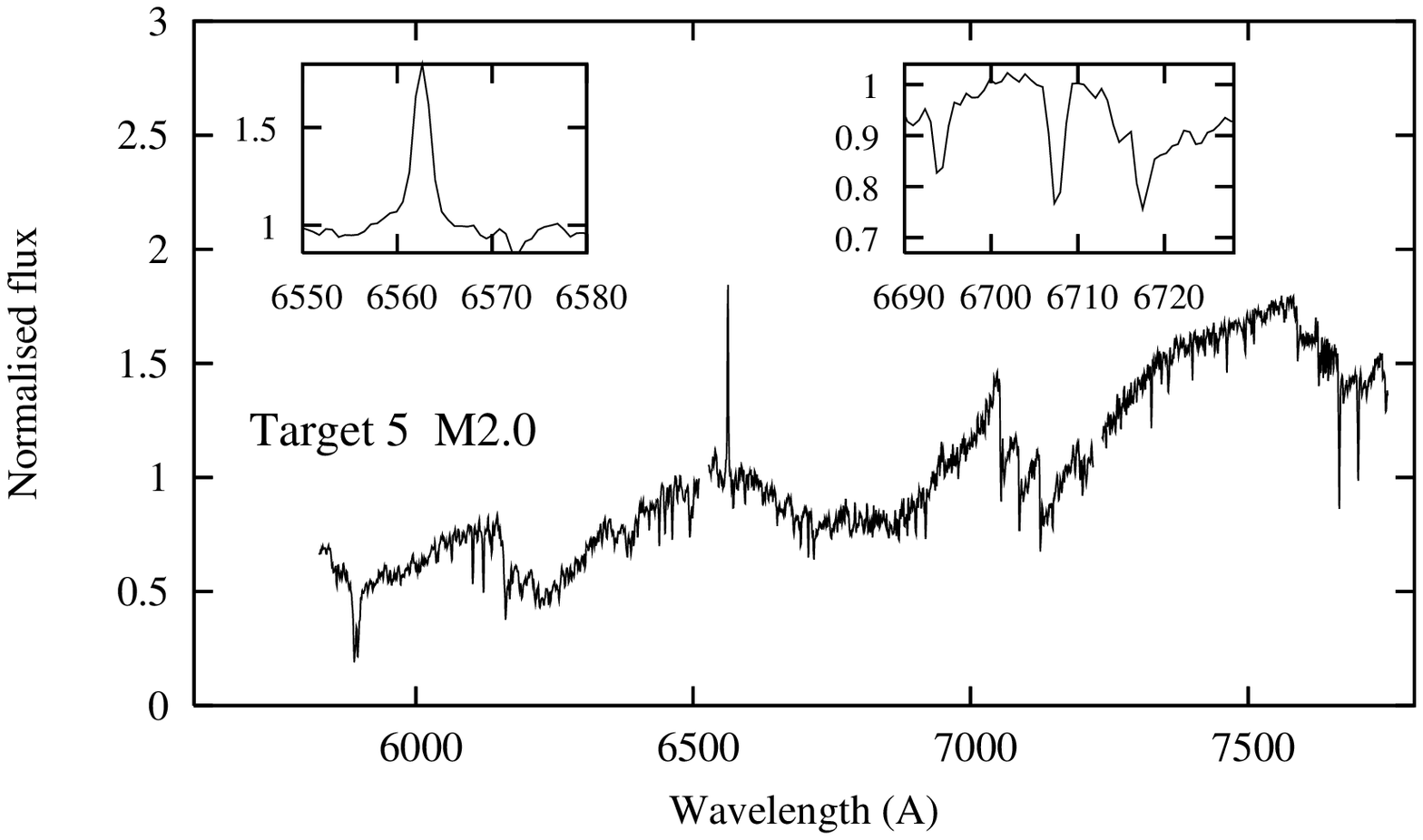}
\includegraphics[width=71mm]{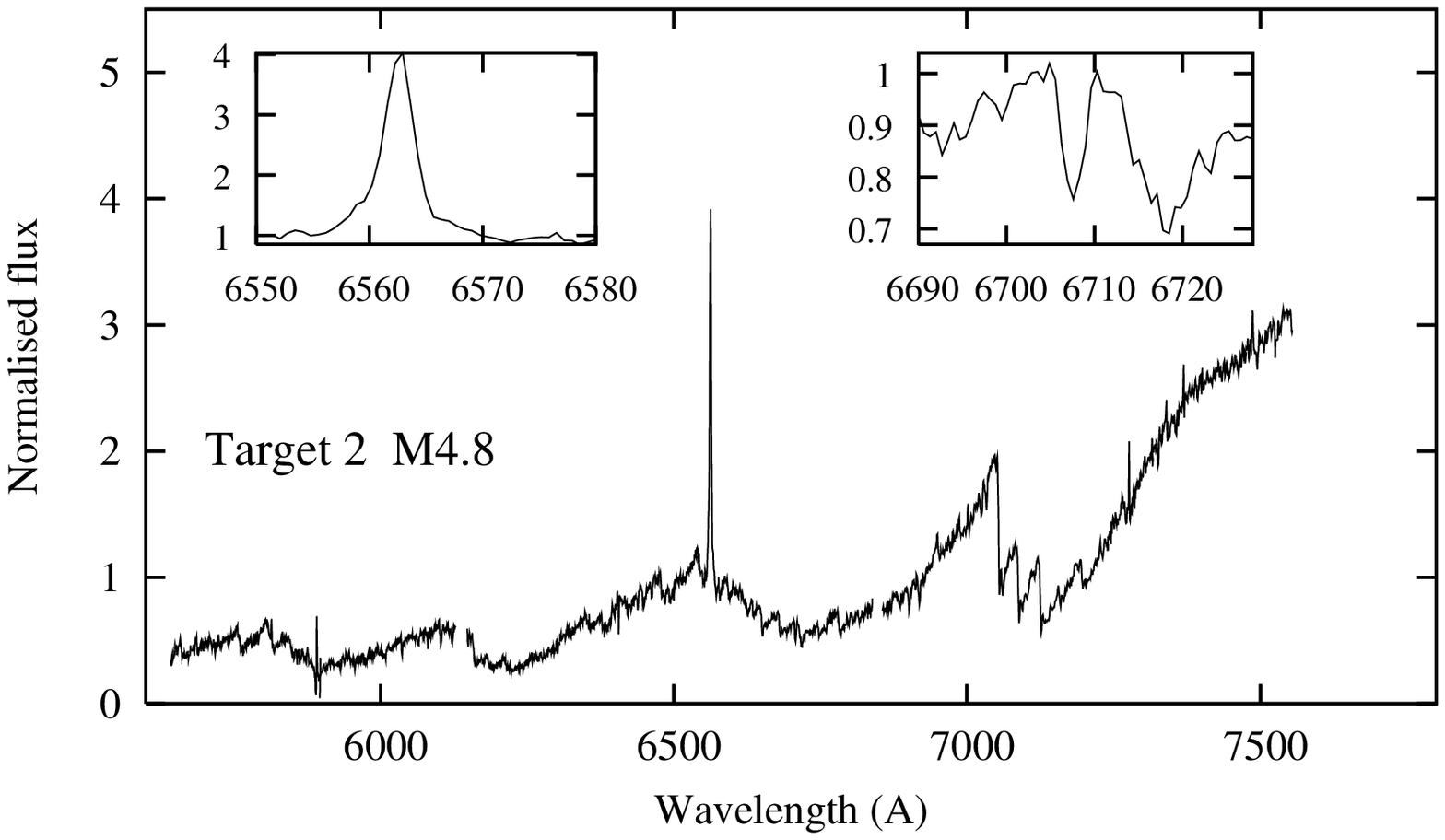}
\end{minipage}
\begin{minipage}[t]{0.45\textwidth}
\includegraphics[width=71mm]{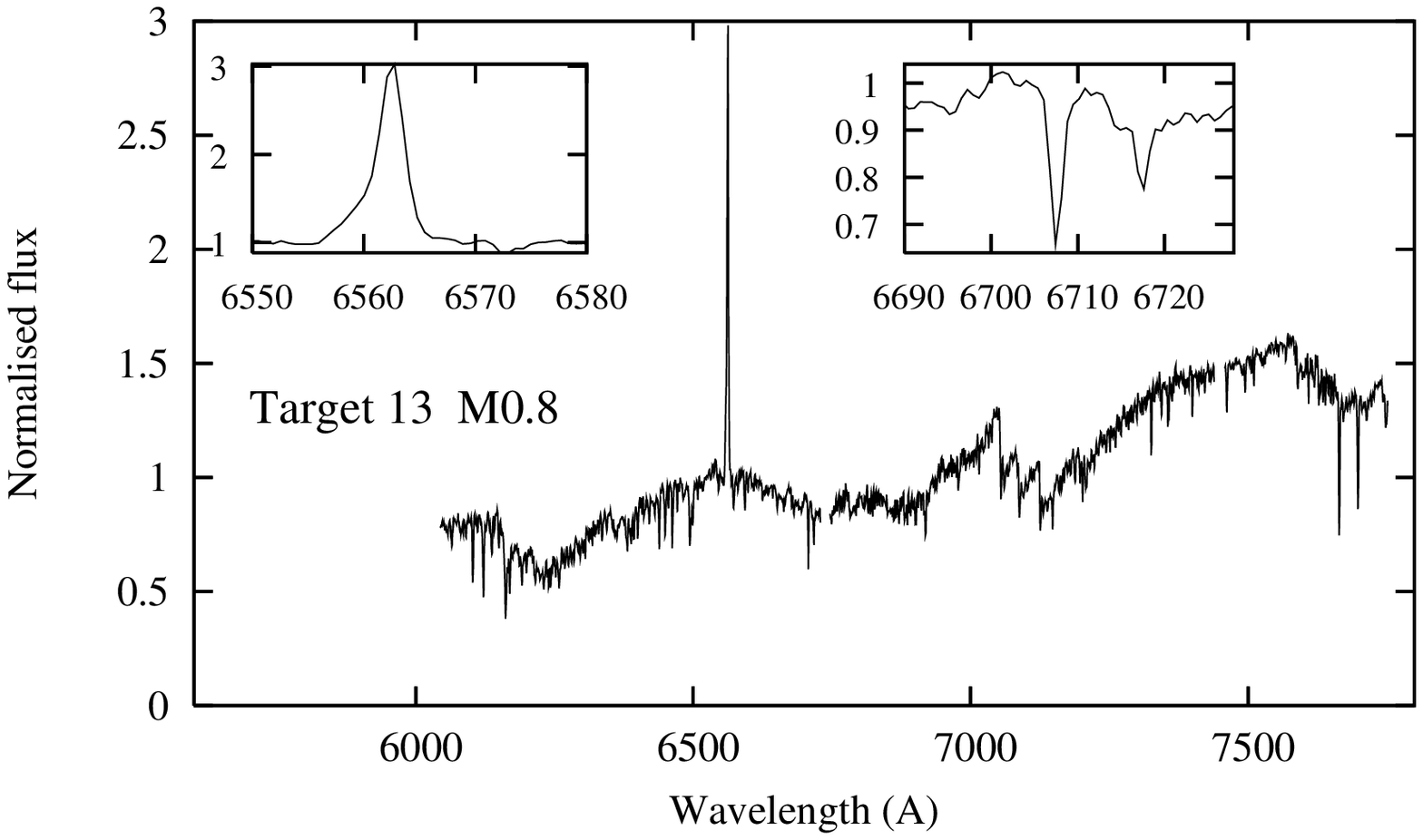}
\includegraphics[width=71mm]{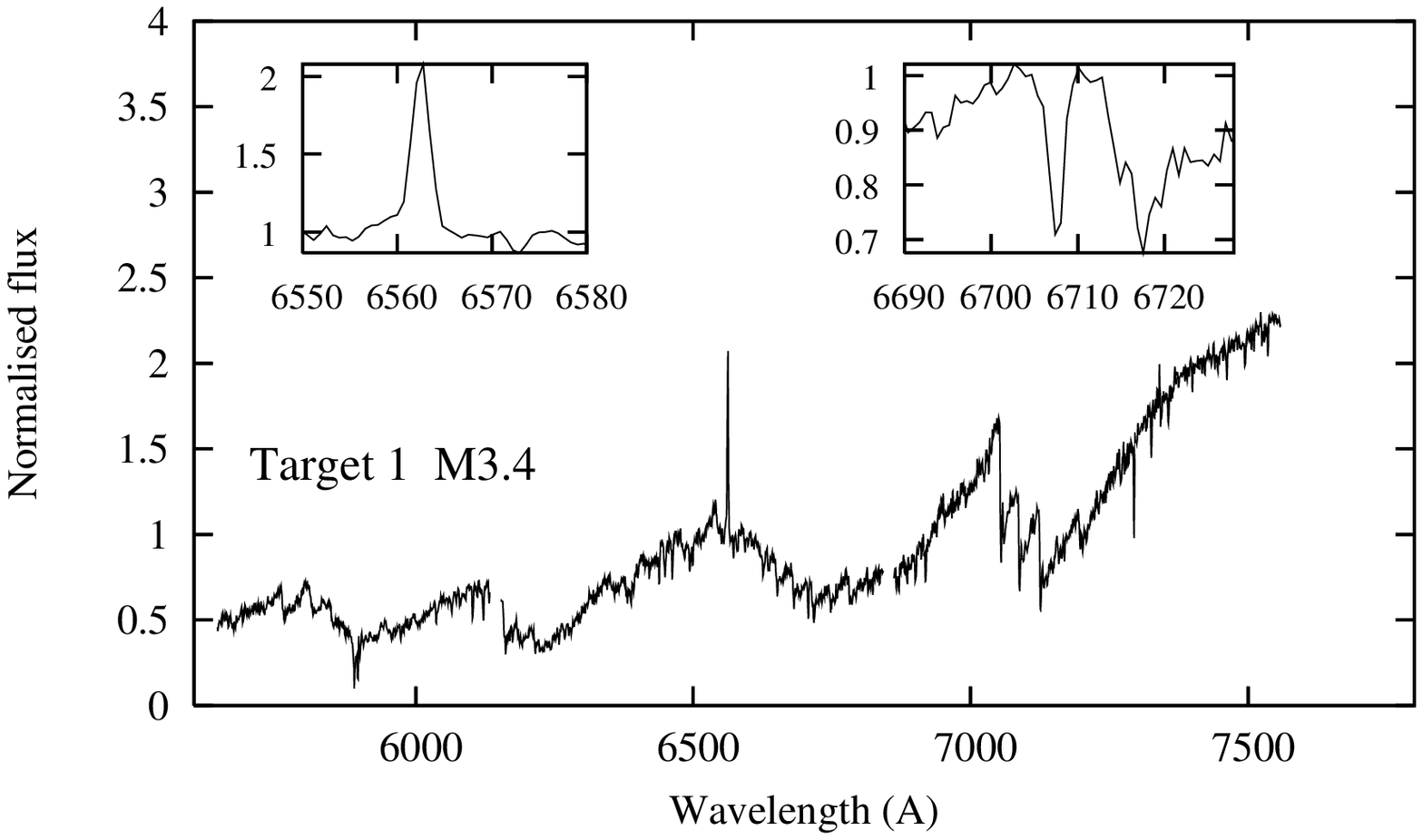}
\includegraphics[width=71mm]{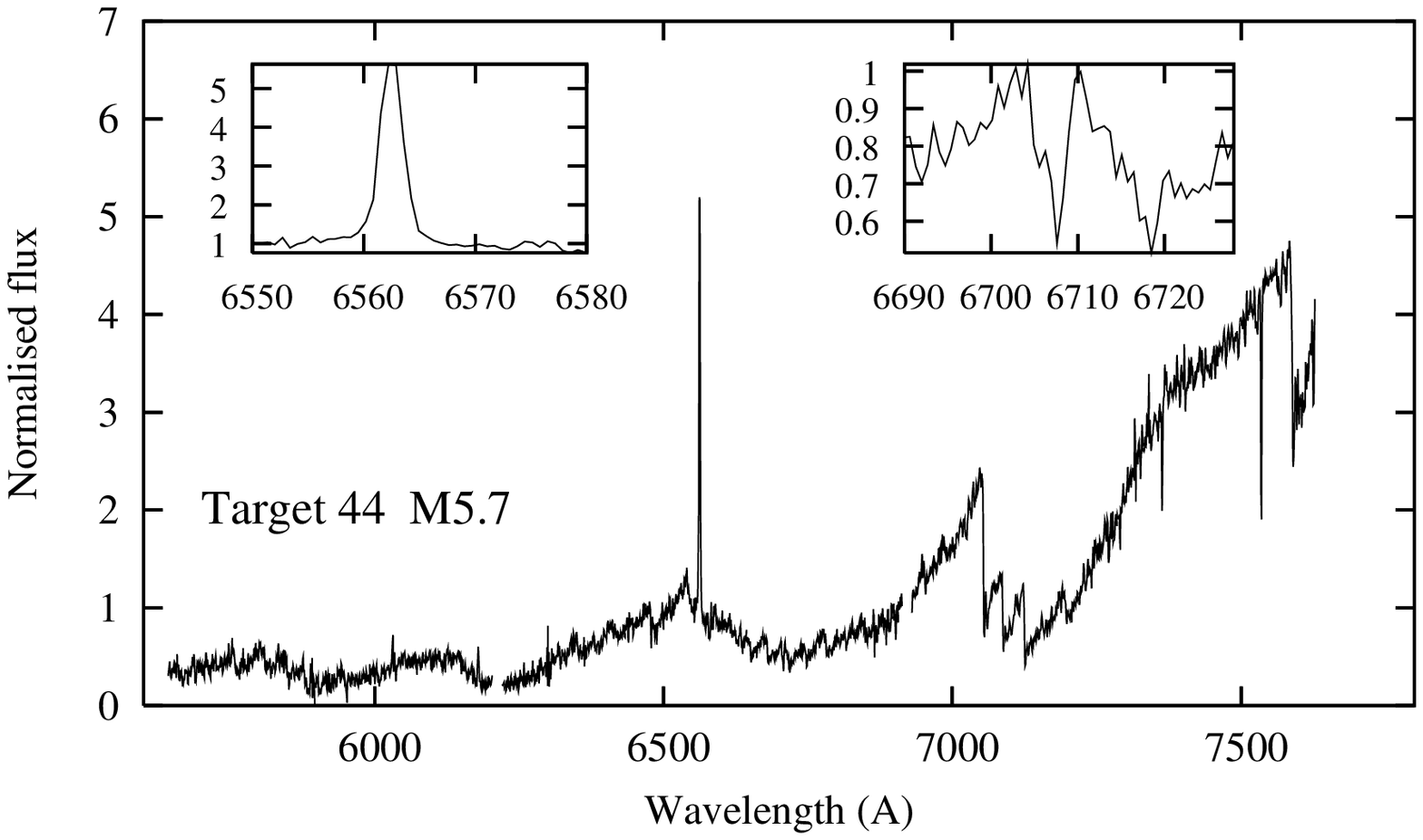}
\end{minipage}
\caption{Example spectra from our target list, covering the full range
  of spectral type and signal-to-noise ratio. Spectra have been subject
  to relative flux calibration, telluric correction and have been
  normalised to a continuum point near H$\alpha$. The inserts on each
  plot show normalised spectra in the regions of the H$\alpha$ and
  Li\,{\sc i}~6708\AA\ lines.
}
\label{specplot}
\end{figure*}

\begin{table*}
\caption{Identification (as given in Table~\ref{ccd_catalogue}),
  positions and photometry for the 47 targets observed in three
  separate GMOS masks.}
\begin{tabular}{crrcccccc}
\hline
Target & \multicolumn{2}{c}{Cat. ID} & RA  & Dec & $I_C$ & $\delta I_C$& $R_C - I_C$ & $\delta (R_C - I_C )$ \\
\hline
\multicolumn{9}{l}{Mask 1}\\
   &    &      &	      &		    &	      &	       &	&\\
01 &  1.04 & 1107 & 06 08 38.879 &+13 54 22.45 &  17.189 &  0.009 &  1.626 &  0.011 \\
02 &  1.04 & 1108 & 06 08 38.959 &+13 56 15.66 &  17.462 &  0.010 &  1.736 &  0.009 \\
03 &  1.04 & 1138 & 06 08 26.094 &+13 54 44.42 &  17.303 &  0.009 &  1.724 &  0.012 \\
04 &  1.04 & 1144 & 06 08 24.267 &+13 55 13.57 &  17.187 &  0.009 &  1.645 &  0.011 \\
05 &  1.04 & 1372 & 06 08 31.196 &+13 57 40.47 &  16.520 &  0.008 &  1.334 &  0.010 \\
06 &  1.04 & 1600 & 06 08 37.510 &+13 57 47.18 &  18.002 &  0.012 &  1.733 &  0.014 \\
07 &  1.04 & 1604 & 06 08 35.994 &+13 56 54.98 &  18.142 &  0.012 &  1.731 &  0.013 \\
08 &  1.04 &  170 & 06 08 41.962 &+13 55 41.30 &  14.605 &  0.008 &  0.575 &  0.011 \\
09 &  1.04 &  186 & 06 08 37.015 &+13 52 43.91 &  15.608 &  0.007 &  0.784 &  0.009 \\
10 &  1.04 & 1924 & 06 08 33.139 &+13 54 59.24 &  18.519 &  0.014 &  1.834 &  0.017 \\
11 &  1.04 &  214 & 06 08 32.212 &+13 53 18.75 &  15.076 &  0.009 &  0.780 &  0.012 \\
12 &  1.04 &  217 & 06 08 32.329 &+13 57 00.56 &  14.893 &  0.009 &  0.773 &  0.011 \\
13 &  1.04 &  252 & 06 08 26.809 &+13 53 37.05 &  16.030 &  0.008 &  1.129 &  0.009 \\
14 &  1.04 & 3039 & 06 08 33.925 &+13 53 47.62 &  19.339 &  0.025 &  2.075 &  0.047 \\
15 &  1.04 &  542 & 06 08 39.554 &+13 57 12.76 &  16.171 &  0.008 &  1.196 &  0.009 \\
16 &  1.04 &  555 & 06 08 34.112 &+13 55 46.67 &  16.313 &  0.008 &  1.497 &  0.009 \\
17 &  1.04 &  900 & 06 08 33.554 &+13 56 24.88 &  17.093 &  0.009 &  1.668 &  0.011 \\
18 &  1.04 &  903 & 06 08 32.401 &+13 55 27.38 &  16.981 &  0.009 &  1.620 &  0.010 \\
   &    &      &	      &		    &	      &	       &	&\\
\multicolumn{9}{l}{Mask 2}\\
   &    &      &	      &		    &	      &	       &	&\\
19 &  1.04 & 1066 & 06 08 50.529 &+14 01 05.12 &  17.340 &  0.009 &  1.624 &  0.011\\ 
20 &  1.04 & 1103 & 06 08 40.244 &+13 59 33.45 &  17.261 &  0.009 &  1.761 &  0.009 \\
21 &  1.04 & 1318 & 06 08 51.862 &+14 02 21.58 &  17.597 &  0.010 &  1.731 &  0.013 \\
22 &  1.04 & 1556 & 06 08 52.321 &+14 01 40.77 &  18.228 &  0.013 &  1.833 &  0.015 \\
23 &  1.04 &  159 & 06 08 43.331 &+14 00 14.38 &  15.973 &  0.007 &  0.967 &  0.008 \\
24 &  1.04 & 1610 & 06 08 34.387 &+13 57 48.18 &  17.487 &  0.010 &  1.797 &  0.015 \\
25 &  1.04 &  168 & 06 08 42.279 &+14 01 11.94 &  15.905 &  0.007 &  1.026 &  0.008 \\
26 &  1.04 &  195 & 06 08 35.413 &+13 58 41.14 &  14.893 &  0.009 &  0.788 &  0.011 \\
27 &  1.04 & 2269 & 06 08 35.533 &+13 58 58.60 &  18.797 &  0.018 &  1.977 &  0.028 \\
28 &  1.04 & 3044 & 06 08 33.238 &+13 58 07.51 &  18.720 &  0.027 &  1.800 &  0.096 \\
29 &  1.04 &  543 & 06 08 39.096 &+14 00 26.21 &  16.270 &  0.008 &  1.300 &  0.009 \\
30 &  1.04 &  552 & 06 08 34.845 &+13 59 13.44 &  16.121 &  0.008 &  1.164 &  0.009 \\
31 &  1.04 &  554 & 06 08 34.489 &+14 01 31.66 &  16.394 &  0.008 &  1.650 &  0.009 \\
32 &  1.04 &  718 & 06 08 33.516 &+13 59 52.69 &  16.842 &  0.008 &  1.450 &  0.010 \\
33 &  1.04 &  875 & 06 08 43.644 &+14 00 44.47 &  16.964 &  0.009 &  1.745 &  0.010 \\
   &    &      &	      &		    &	      &	       &	&\\
\multicolumn{9}{l}{Mask 3}\\
   &    &      &	      &		    &	      &	       &	&\\
28 &  1.04 & 3044 & \multicolumn{6}{l}{repeated in this mask} \\
34 &  1.04 &  906 & 06 08 28.739 &+13 58 27.40 &  16.481 &  0.012 &  1.337 &  0.012\\ 
35 &  1.04 & 1154 & 06 08 19.214 &+13 59 30.35 &  15.728 &  0.007 &  0.826 &  0.011 \\
36 &  1.04 & 1367 & 06 08 33.076 &+14 01 46.38 &  17.817 &  0.010 &  1.763 &  0.011 \\
37 &  1.04 & 1622 & 06 08 32.337 &+13 59 03.87 &  17.755 &  0.012 &  1.775 &  0.013 \\
38 &  1.04 & 1641 & 06 08 26.254 &+13 57 45.79 &  17.665 &  0.010 &  1.789 &  0.011 \\
39 &  1.04 & 1947 & 06 08 27.496 &+13 58 49.92 &  18.230 &  0.013 &  1.869 &  0.017 \\
40 &  1.04 &  202 & 06 08 34.120 &+14 00 24.11 &  15.955 &  0.008 &  1.465 &  0.009 \\
41 &  1.04 &  224 & 06 08 31.387 &+14 01 38.91 &  15.730 &  0.007 &  1.302 &  0.008 \\
42 &  1.04 & 2349 & 06 08 17.379 &+13 59 55.62 &  16.433 &  0.032 &  1.609 &  0.059 \\
43 &  1.04 &  260 & 06 08 25.478 &+14 02 11.90 &  14.866 &  0.009 &  1.045 &  0.011 \\
44 &  1.04 & 2684 & 06 08 27.158 &+13 57 29.07 &  18.685 &  0.017 &  2.037 &  0.029 \\
45 &  1.04 &  563 & 06 08 28.518 &+13 59 45.31 &  16.165 &  0.008 &  1.239 &  0.009 \\
46 &  1.04 &  747 & 06 08 21.831 &+13 57 58.72 &  16.551 &  0.008 &  1.413 &  0.009 \\
47 &  1.04 &  929 & 06 08 20.784 &+14 00 56.85 &  16.609 &  0.008 &  1.457 &  0.010 \\
\hline
\end{tabular}
\label{targetlist}
\end{table*}

The Gemini Multi-Object Spectrograph (GMOS) was used in conjunction
with slit masks at the Gemini North telescope 
to observe 47 candidate low-mass members of NGC~2169
with $14.6<I_C<19.3$. This corresponds to a mass range (for an assumed
distance of 1000~pc and age of 10\,Myr) of $0.14<M/M_{\odot}<1.3$
according to the models of Baraffe et al. (1998). Stars were targeted
based on their location in the colour-magnitude diagram and with the
aim of maximising the number of targets that could be included in three
separate slit mask designs (see Table~\ref{slitmask} and Fig.~\ref{gaiaplot}).
Table~\ref{targetlist} gives the coordinates and photometry of the
targets, which are split according to which of the three masks they
were observed in. One target (number 28) was observed in both masks 2 and 3.

We used slits of width 0.5 arcsec and with lengths of approximately
8--10 arcsecs. The R831 grating was used with a long-pass OG515 filter
to block second order contamination. The resolving power was 4400 and
simultaneous sky subtraction of the spectra was possible. The spectra
covered $\sim 2000$\AA, with a central wavelength of 6200\AA --
7200\AA\ depending on the location of the slits within the 5.5 arcminute
field of view.

Observations were taken in queue mode (program number GN-2005B-Q-30)
through three separate masks during October and December 2005 (see
Table~\ref{slitmask}). For each mask we obtained $3\times1800$\,s
exposures bracketed by observations of a CuAr lamp for wavelength
calibration and a quartz lamp for flat-fielding and slit location.  The
spectra were recorded on three $2048\times4068$ EEV chips leading to
two $\simeq 16$\AA\ gaps in the coverage. The CCD pixels were binned
$2\times2$ before readout, corresponding to $\sim 0.67$\AA\ per binned
pixel in the dispersion direction and 0.14 arcsec per binned pixel in
the spatial direction. Conditions were clear with seeing of 0.5--0.6
arcsec (FWHM measured from the spectra).

The data were reduced using version 1.8.1 of the GMOS data reduction
tasks running with version 2.12.2a of the Image Reduction and Analysis
Facility (IRAF). The data were bias subtracted, mosaiced and
flat-fielded. A two-dimensional wavelength calibration solution was
provided by the arc spectra and then the target spectra were
sky-subtracted and extracted using 2 arcsec apertures. The three
individual spectra for each target were then combined using a rejection
scheme which removed obvious cosmic rays.  The instrumental
wavelength response was removed from the combined spectra using
observations of a white dwarf standard to provide a relative flux
calibration. The same calibration spectrum was used to construct a
telluric correction spectrum. A scaled version of this was divided into
the target spectra, tuned to minimise the RMS in regions dominated by
telluric features.

The SNR of each combined extracted
spectrum was estimated empirically from the RMS deviations of straight
line fits to segments of ``pseudo-continuum'' close to the
\lii~6708\AA\ features (see below). As small unresolved spectral
features are expected to be part of these pseudo-continuum regions,
these SNR estimates, which range from $\sim 10$--20 in the faintest
targets to $>200$ in the brightest, should be lower limits. Examples of
the reduced spectra are shown in Fig.~\ref{specplot}. All the reduced
spectra are available in ``fits'' format from the ``Cluster'' Collaboration's home page
(see footnote 1).

\subsection{Analysis}
\label{analysis}
Each spectrum was analysed to yield a spectral type, equivalent widths
of the \lii~6708\AA\ and H$\alpha$ lines and a heliocentric RV.
Each of these analyses is described below. The results are given in
Table~\ref{specresults}.

\begin{table*}
\caption{Results from the spectral analysis of section~\ref{analysis},
  listing spectral type, estimated signal-to-noise ratio, equivalent
  widths of the Li\,{\sc i}~6708\AA\ and H$\alpha$ lines and the heliocentric
  radial velocity. The last column gives the overall membership
  assessment -- Y  member, N  non-member, ?  membership questionable
  (see section \ref{member}).}
\begin{tabular}{cccccccccc}
\hline
Target & Spectral  & SNR & EW(Li)  & $\delta$EW(Li) & FWHM(Li) &
EW(H$\alpha$) & RV & $\delta RV$ & Mem? \\
       & Type      & & (\AA)   & (\AA)          & (\AA)    &
(\AA)         &	(\kms) & (\kms)   \\
\hline
\multicolumn{10}{l}{Mask 1}\\
&   &    &      &	      &		    &	      &	       &	&\\
01 &   M3.4 & 77  &  0.62  & 0.02 & 1.86  &  -3.0  & 11.7  &  1.8 & Y\\
02 &   M4.8 & 41  &  0.68  & 0.05 & 2.54  & -12.1  & 16.6  &  5.5 & Y\\
03 &   M4.2 & 45  &  0.50  & 0.04 & 1.86  &  -4.3  & 14.4  &  2.8 & Y\\
04 &   M3.7 & 95  &  0.52  & 0.02 & 1.74  &  -3.7  & 15.1  &  2.2 & Y\\
05 &   M2.0 &100  &  0.45  & 0.02 & 1.65  &  -2.1  & 11.7  &  0.2 & Y\\
06 &   M4.7 & 37  &  0.63  & 0.04 & 1.97  &  -5.4  & 12.6  &  3.2 & Y\\
07 &   M4.4 & 23  &$<0.24$ & --   &  --   &  -0.1  &-33.0  &  3.6 & N\\
08 &   K5.3 &180  &$<0.03$ & --   &  --   &  -1.7  & 29.9  & 16.7 & ?\\
09 &   K5.6 &220  &$<0.03$ & --   &  --   &  +0.5  &-27.2  &  5.0 & N\\
10 &   M5.3 & 16  &  0.67  & 0.09 & 1.57  &  -5.7  & 14.7  &  3.8 & Y\\
11 &   K5.7 &156  &  0.47  & 0.01 & 1.53  &  -0.9  & 24.5  &  8.8 & Y\\
12 &   K5.7 &212  &  0.52  & 0.01 & 2.35  &  -1.3  & 19.4  &  6.1 & Y\\
13 &   M0.8 &108  &  0.59  & 0.01 & 1.62  &  -6.6  &  8.5  &  1.7 & Y\\
14 &   M5.5 &  8  &$<0.68$ & --   &  --   & -18.8  & 18.3  &  3.8 & ?\\
15 &   M1.1 &133  &  0.52  & 0.01 & 1.62  &  -2.5  &  8.7  &  2.0 & Y\\
16 &   M2.8 & 76  &  0.56  & 0.02 & 1.79  &  -8.9  & 13.7  &  2.2 & Y\\
17 &   M3.5 & 67  &  0.54  & 0.02 & 1.81  &  -3.6  & 14.9  &  2.4 & Y\\
18 &   M3.7 & 52  &  0.63  & 0.03 & 1.74  &  -3.3  & 16.4  &  1.7 & Y\\
&   &    &      &	      &		    &	      &	       &	&\\
\multicolumn{10}{l}{Mask 2}\\
&   &    &      &	      &		    &	      &	       &	&\\
19 &   M3.4 & 86  &  0.49  & 0.02 & 1.86  &  -1.1  & 14.1  &  5.7 & Y\\
20 &   M4.9 & 48  &  0.61  & 0.03 & 1.76  &  -1.2  & 20.1  &  3.3 & Y\\
21 &   M4.2 & 27  &$<0.20$ & --   &  --   &  --   & 27.9  &  3.9 & N\\
22 &   M5.0 & 32  &  0.84  & 0.06 & 3.12  &  -9.4  & 16.3  &  8.5 & Y\\
23 &   K7.3 &111  &$<0.05$ & --   &  --   &  +0.9  &-36.5  &  3.4 & N\\
24 &   M4.2 & 40  &  0.64  & 0.04 & 2.26  &  -4.8  & 14.4  &  2.6 & Y\\
25 &   K7.7 &173  &  0.53  & 0.01 & 1.69  &  -1.7  &  9.3  &  3.3 & Y\\
26 &   K5.9 &266  &  0.47  & 0.01 & 1.41  &  -1.3  & 12.1  &  3.5 & Y\\
27 &   M5.6 & 20  &  0.71  & 0.07 & 1.34  & -13.6  & 25.0  &  6.4 & Y\\
28 &   M5.3 & 14  &  0.98  & 0.14 & 2.84  & -10.4  & 20.4  &  4.2 & Y\\
29 &   M1.4 & 78  &  0.57  & 0.02 & 1.71  &  -6.4  & 10.5  &  1.7 & Y\\
30 &   M0.5 & 80  &  --    & --   & 1.97  &  -3.5  &  8.4  &  1.8 & ?\\
31 &   M3.7 & 53  &  0.54  & 0.03 & 1.81  &  -6.7  & 13.4  &  2.3 & Y\\
32 &   M3.0 & 61  &  0.37  & 0.02 & 1.67  &  -3.5  & 13.4  &  1.4 & Y\\
33 &   M4.0 & 39  &  0.56  & 0.04 & 1.65  &  -5.9  & 12.8  &  2.7 & Y\\
&   &    &      &	      &		    &	      &	       &	&\\
\multicolumn{10}{l}{Mask 3}\\
&   &    &      &	      &		    &	      &	       &	&\\
28 &   M5.3 & 20  &  1.07  & 0.09 & 2.32  &  -7.9  & 17.8  &  8.1 & Y\\
34 &   M2.2 & 75  &  0.54  & 0.02 & 1.65  &  -3.8  & 12.1  &  2.0 & Y\\
35 &   K5.9 &126  &$<0.05$ & --   &  --   &  +0.6  &-53.3  &  2.3 & N\\
36 &   M4.9 & 55  &  0.59  & 0.03 & 1.69  &  -5.7  & 15.9  &  5.9 & Y\\
37 &   M5.0 & 41  &  0.71  & 0.04 & 1.81  &  -7.3  & 18.7  &  5.3 & Y\\
38 &   M4.7 & 25  &  0.61  & 0.06 & 1.79  &  -6.6  & 18.8  &  3.5 & Y\\
39 &   M5.0 & 35  &  0.63  & 0.06 & 2.84  &  -6.2  & 21.4  &  4.5 & Y\\
40 &   M3.0 & 92  &  0.62  & 0.02 & 2.02  &   --   &  6.4  &  5.3 & Y\\
41 &   M1.7 & 87  &  --    & --   & --    &  -4.4  &  4.0  &  3.4 & ?\\
42 &   M3.8 & 46  &  --    & --   & --    &  -5.8  & 26.7  &  2.4 & ?\\
43 &   K5.1 &134  &$<0.04$ & --   & --    &  +1.5  & 27.4  &  6.9 & N\\
44 &   M5.7 & 17  &  1.10  & 0.12 & 3.17  & -11.7  & 18.3  &  4.5 & Y\\
45 &   M1.8 & 76  &  0.59  & 0.02 & 1.57  &  -3.7  & 12.4  &  1.9 & Y\\
46 &   M2.6 &123  &  0.61  & 0.01 & 1.81  &  -2.6  & 17.5  &  1.4 & Y\\
47 &   M1.8 & 61  &  0.58  & 0.03 & 1.72  &  -2.3  & 16.6  &  0.9 & Y\\
\hline
\end{tabular}
\label{specresults}
\end{table*}

\subsubsection{Spectral Types}
\label{spt}

\begin{figure}
\includegraphics[width=75mm]{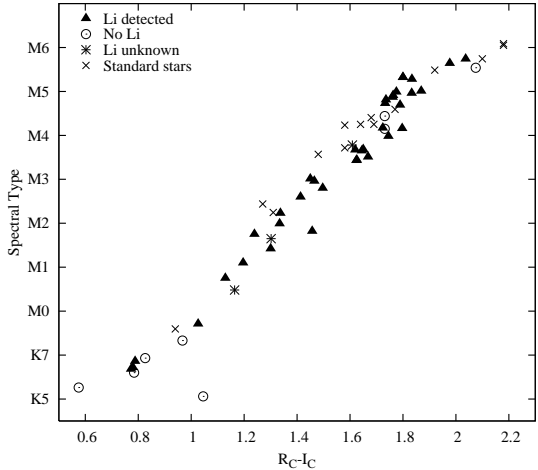}
\caption{Spectral types deduced from the TiO(7140\AA)
  index as a function of $R_C-I_C$. 
  Also shown are results from the spectra
  of standard stars where $R_C-I_C$ is available (see text).}
\label{sptri}
\end{figure}

\begin{table}
\caption{The relationship between TiO(7140\AA) index and spectral type}
\begin{tabular}{cc}
\hline
Index & Spectral Type \\
\hline
0.99 & K5 \\
1.13 & K7 \\
1.26 & M0 \\
1.40 & M1 \\
1.53 & M2 \\
1.74 & M3 \\
2.08 & M4 \\
2.61 & M5 \\
3.38 & M6 \\
\hline
\end{tabular}
\label{sptindex}
\end{table}

Spectral types were estimated from the strength of
the TiO(7140\AA) narrow band spectral index (see Oliveira et
al. 2003). This index is quite temperature sensitive and can be
calibrated for spectral type using spectra of well
known late K and M-type field dwarfs. We constructed a polynomial
relationship between spectral type and the TiO(7140\AA) index that was
used to estimate the spectral type of our targets, based on a
numerical scheme where M0--M6$=$0--6, K5\,$=-2$ and
K7\,$=-1$. Table~\ref{sptindex} gives the adopted relationship between
the TiO(7140\AA) index and spectral type. The scatter
around the polynomial indicates that these spectral types are good to
$\pm$ half a subclass for stars of type M0 and later, but to
only a subclass at earlier type stars as the molecular band is
very weak at spectral type K5.

A plot of spectral type derived from the TiO(7140\AA) index versus
$R_C-I_C$ colour reveals a smooth relationship (see
Fig.~\ref{sptri}) with little scatter.
We expect this even if there is some contamination by field
interlopers in the sample. These would have similar colours and
spectral types to the cluster members, but would be foreground objects
with lower luminosities. A comparison of the positions of standard
stars on this plot ($R_C-I_C$ colours where available are from Leggett
1992) reveals an average redward offset of $\simeq 0.05$ mag in the $R_C-I_C$
values of our targets at a given spectral type. Of course we
expect cluster members to have suffered a reddening $E(R_C-I_C)\simeq 0.14$ mag
(corresponding to $E(B-V)=0.20$ mag -- Taylor 1986). This
comparison demonstrates that the photometric calibration for these
red stars is reasonable, although there is a hint that the $R_C-I_C$
values may be too blue at the reddest colours. We have to temper this conclusion
with the probability that the relationship between colour and spectral
type is slightly different in very young stars.

\subsubsection{Lithium measurements}
\label{li}

The \lii~6708\AA\ resonance feature should be strong in cool young stars
with undepleted Li -- with an equivalent width (EW) of 0.5\AA\ to
0.6\AA\ according to the curves of growth presented by Zapatero-Osorio
et al. (2002). Insets in Fig.~\ref{specplot} show the Li region in
a number of our targets. Lithium is an ephemeral element in the
atmospheres of very cool stars, and its presence in the photospheres of
late K and M-type stars is a strong indicator
of youth.

The EW of the \lii~6708\AA\ feature was estimated by fitting it with a
Gaussian function. We preferred this to direct integration because in
lower signal-to-noise spectra we eliminate the subjectivity involved in
choosing the integration limits and we get a straightforward indication
of rapid rotation (see below).  The ``pseudo-continuum'' was
estimated using straight line fits to the regions immediately around
the Li feature, excluding regions beyond 6712\AA\ which contain a
strong Ca line and which are noisy due to the subtraction of a strong
S\,{\sc ii} sky line. None of the Li lines in our spectra show any strong evidence
for a non-Gaussian shape or double lines; target 44 (shown in
Fig.~\ref{specplot}) has the most ``non-Gaussian'' appearance, but even
here the Gaussian fit is only rejected at 93 per cent confidence and in
any case the EW estimated by direct integration would not differ
significantly from the Gaussian result. The EW and Gaussian FWHM of the
Li lines are given in
Table~\ref{specresults}. Uncertainties in the EW are estimated using
the formula $\delta\,{\rm EW} = \sqrt{2fp}/{\rm SNR}$, where $2f$ is
twice the Gaussian FWHM of the line (approximately the range over which
the EW is integrated) and $p$ is the pixel size
(0.67\AA).\footnote{This formula arises from adding the
  uncertainties in each pixel flux estimate in quadrature, assuming
  that these uncertainties are given by the average signal-to-noise
  ratio and
  that the line is integrated over a range of $2f$. The additional statistical
  uncertainty due to the continuum level estimate is small in
  comparison.} In 8 cases
there was no obvious Li feature to measure, in which case a $3\sigma$
upper limit is quoted. In 3 cases the Li feature fell in a gap between
the detectors and no EW could be measured.

For the majority of the sample there are clear detections of the Li
feature with EW\,$>0.3$\AA. Comparisons with Li-depletion
patterns in open clusters of known age (e.g. see Fig.~10 of Jeffries et
al. 2003) place empirical age constraints on these
stars. Li EWs of $>0.3$\AA\ are not seen for any stars of spectral type
cooler than K5 in the Pleiades or Alpha Per clusters (with ages of
120\,Myr and 90\,Myr respectively and excepting the very low
luminosity stars beyond the lithium depletion boundary where Li remains
unburned -- see Soderblom et al. 1993; Jones et al. 1996).  Nor can
strong Li lines be seen in M dwarfs of the 35--55\,Myr open clusters
NGC~2547, IC~2391 and IC~2602 (see Randich et al. 2001; Jeffries et
al. 2003; Barrado y Navascu\'es et al. 2004; Jeffries \& Oliveira 2005,
and again excepting the very cool M dwarfs beyond the lithium depletion
boundary).  In summary we assume that 
objects with EW[Li]\,$>0.3$\AA\ are all younger than 100\,Myr
and younger than 50\,Myr if they have spectral type $\geq$ M0. These
Li-rich objects are therefore very likely to be members of NGC 2169 and this
conclusion is supported by RV measurements (see
below). However the converse may not be true -- a lack of Li is not
used as the sole criterion for excluding a candidate member, as one of
the aims of this paper is to investigate possible instances of
anomalously large Li depletion.

In a number of cases the FWHM of the \lii~6708\AA\ line is
significantly broader than the 1.7\AA\ expected from the intrinsic
width of the doublet and the instrumental resolution. In these cases 
rotational broadening is suspected, which implies projected rotational
velocities from 25\kms\ (for a FWHM of 2\AA) up to about 60\kms\ for
the broadest lines.

\begin{figure}
\includegraphics[width=75mm]{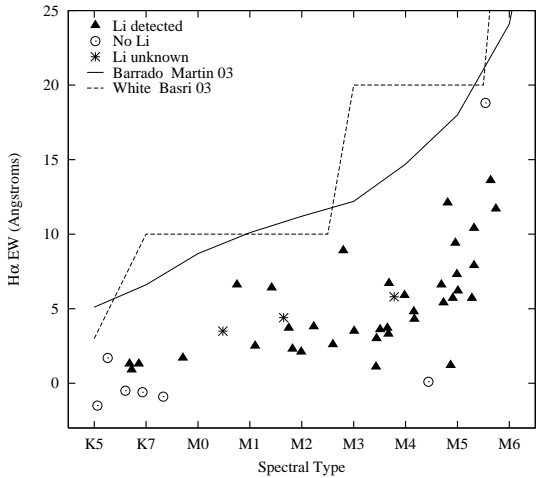}
\caption{H$\alpha$ EW as a function of spectral type. Objects with Li,
  without Li or where the Li status is unknown are indicated. Two lines are
  shown that have previously been used to separate accretion-generated
  H$\alpha$ emission from a lower level of emission that could be
  attributable to a chromosphere (Barrado y Navascu\'es \& Mart\'in
  2003; White \& Basri 2003).
  }
\label{haspt}
\end{figure}

\subsubsection{H$\alpha$ measurements and circumstellar material}

\label{accrete}

\begin{figure*}
\centering
\begin{minipage}[t]{0.45\textwidth}
\includegraphics[width=71mm]{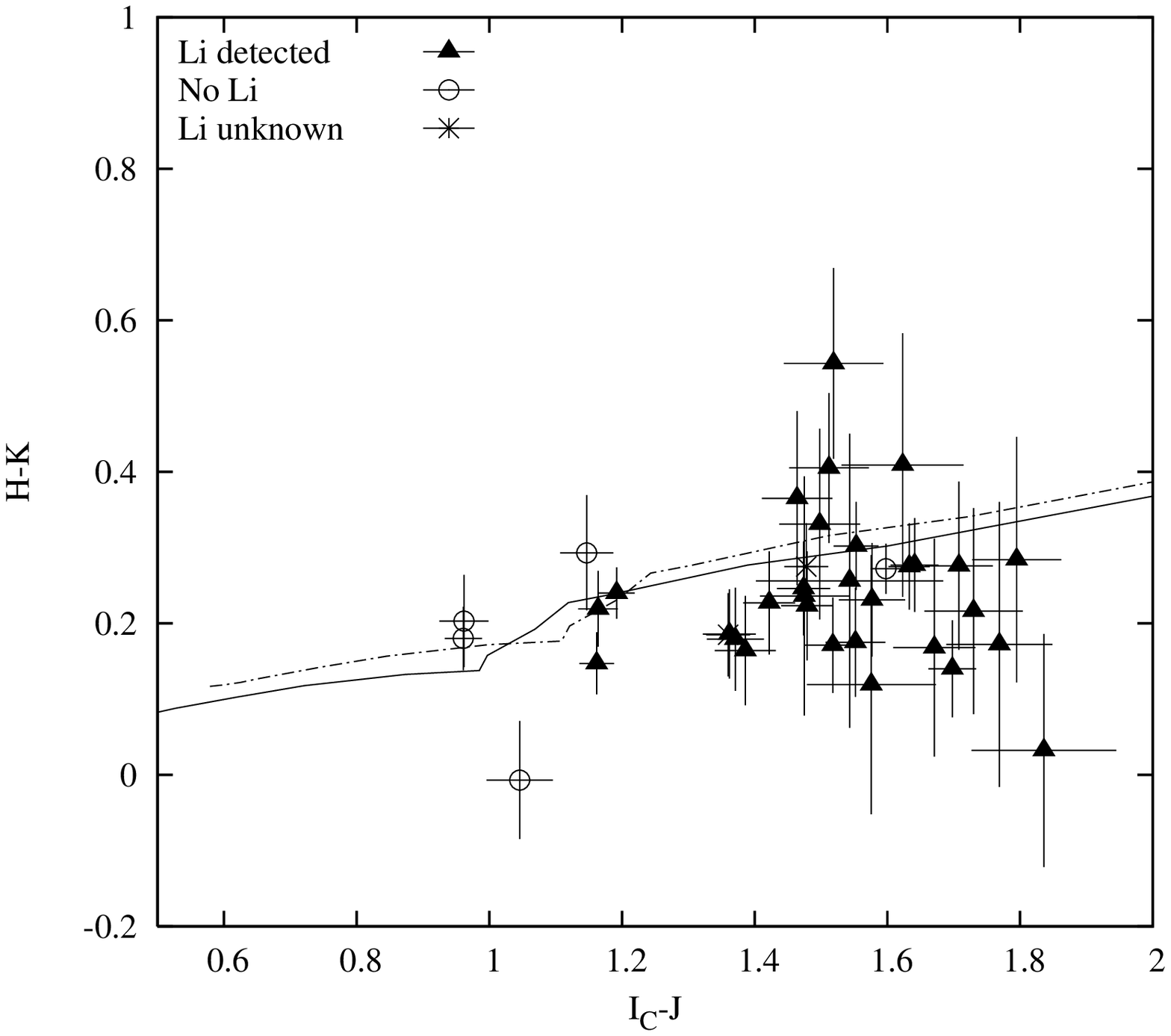}
\end{minipage}
\begin{minipage}[t]{0.45\textwidth}
\includegraphics[width=71mm]{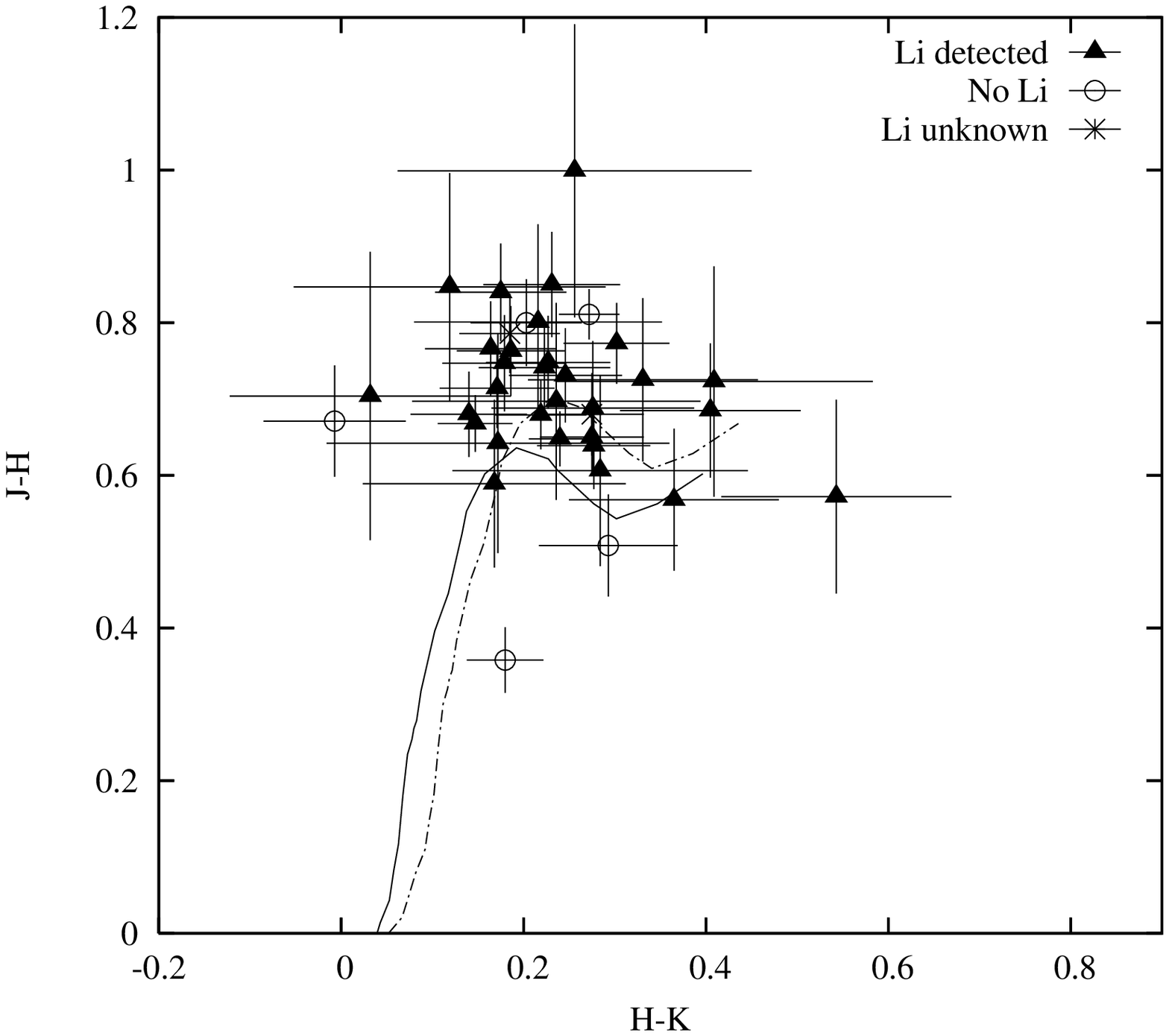}
\end{minipage}
\caption{Near infrared colour-colour diagrams 
 for candidate members of NGC 2169. Objects with Li,
  without Li or where the Li status is unknown are indicated. The data
 plotted are our own $I_C$ magnitudes combined with $JHK$ magnitudes from
 2MASS (Cutri et al. 2003). Only points with uncertainties lower than
 0.2 mag in each colour are shown. Solid 
 lines show the intrinsic loci of man sequence dwarfs (from Bessell \&
 Brett 1988). Dashed lines show the effects of reddening corresponding
 to $E(B-V)=0.20$, using the reddening law of Rieke \& Lebofsky (1985).}
\label{ijhk}
\end{figure*}

H$\alpha$ EWs were
measured for all our targets (except two where the feature fell in a
detector gap) by direct integration above (or below) a pseudo-continuum.
The main uncertainty here is the definition of the pseudo-continuum as a
function of spectral type and probably results in uncertainties of order
0.2\AA, even for the bright targets.

H$\alpha$ emission is ubiquitous from young stars. It arises either as
a consequence of chromospheric activity or is generated by accretion
activity in very young objects (e.g. Muzerolle, Calvet \& Hartmann 1998). The
H$\alpha$ emission from accreting ``classical'' T-Tauri stars (CTTS) is
systematically stronger and broader (velocity widths $>270$\kms --
White \& Basri 2003) than the weak line T-Tauri stars (WTTS) where
the emission is predominantly chromospheric. 

An empirical division between CTTS and WTTS can be made either on the
basis of H$\alpha$ EW or the width of the H$\alpha$ emission line
(e.g. Barrado y Navascu\'{e}s \& Mart\'{i}n 2003; White \& Basri
2003). Fig.~\ref{haspt} shows the H$\alpha$ EW of our targets (where
available) versus spectral type along with the empirical dividing line
between CTTS and WTTS defined by Barrado y Navascu\'{e}s \& Mart\'in
and by White \& Basri.  On the basis of these plots none of our targets
are clear examples of CTTS. A caveat here is that we have a single
epoch spectrum. Accretion or chromospheric activity can be variable
phenomena and multiple observations are preferable for a secure
classification. Large variations in H$\alpha$ strength have been seen
in some PMS stars (e.g. Littlefair et al. 2004), however in a recent
paper by Jayawardhana et al. (2006), multiple H$\alpha$ spectra did not
reveal variability that would change the classification of a
significant fraction of young objects. The most likely error would be a
chromospheric flare leading to a CTTS classification for a WTTS.

The profiles of the H$\alpha$ line were inspected for all targets and
apart from targets 8 and 13, none show evidence for velocity widths (at
10 per cent of maximum) in excess of 300\kms. Target 8 is discussed in
Section~\ref{member}, it does not show a \lii~6708\AA\ line and is
probably not a cluster member.  Target 13 (shown in
Fig.~\ref{specplot}) has a narrow, strong \lii\ line and shows a blue
H$\alpha$ emission wing extending to $\sim 200$\kms. The H$\alpha$ EW
lies below the accretion thresholds in Fig.~\ref{haspt}. It is possible
that either accretion at a low level or chromospheric flaring could
explain this observation.  There are a number of other objects with
widths at about the 270\kms\ threshold advocated by White \& Basri
(2003) as an accretion discriminator. However, we note that our
spectral resolution (70\kms\ FWHM) is relatively poor compared with
that used by White \& Basri. That and the fact that some objects appear
to have rotationally broadened photospheric Li lines, mean that this
threshold should be raised considerably in some cases.  None of the
targets show any evidence for other emission lines that are often (but
not always) associated with young accreting low-mass stars, such as
He\,{\sc i}~5876, 6678\AA.

In addition we have checked for any emission from warm circumstellar
dust by plotting the $H-K$ versus $I_C-J$ colour-colour diagram (see
Fig.~\ref{ijhk}a).  $JHK$ photometry for our targets was taken from the
2MASS point source catalogue (Cutri et al. 2003). 
An excess would show up as an anomalously large
$H-K$ colour with respect to the photospheres of dwarf stars with the
same $I_C-J$, although $H-K$ is nowhere near as sensitive to warm dust as
excesses at longer wavelengths.  None of our targets are
significantly($>2$ sigma) discrepant from the main sequence dwarf locus
reddened according to $E(B-V)=0.2$. The same is true of the more
conventional $J-H$ versus $H-K$ diagram (Fig.~\ref{ijhk}b).  In summary
we find no strong evidence for accretion or warm 
circumstellar dust in any of our targets.

\subsubsection{Radial velocities}

\begin{figure}
\includegraphics[width=75mm]{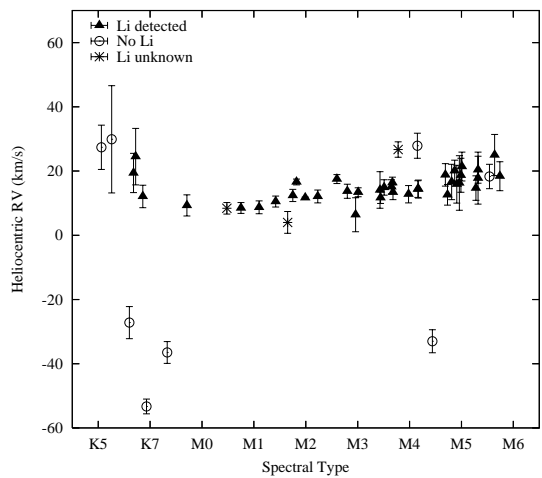}
\caption{Radial velocities as a function of
  spectral type. Stars with and without Li and those with unknown Li
  status are indicated.}
\label{rvspt}
\end{figure}

Our observations were not optimised for measuring RVs -- 
minimal arc calibrations were performed and no RV
standards were observed. Nevertheless we have been able to estimate
RVs of all the targets relative to {\it one} of the
targets and then estimate a zeropoint based on archival spectra.

Relative RVs were determined by cross-correlation against target
5. We chose this star to act as a template because it has a high SNR
and with a spectral type of M2, it has spectral features in common with
targets of both earlier and later spectral types.

A correlation wavelength range of 6600--7400\AA\ was used for stars of
spectral type M0 and later. For earlier spectral types where molecular
features become small then the range 6000--6500\AA\ was used. Raw
correlation lags were adjusted to the same heliocentric reference frame. A
further correction to the RV zero point was estimated by
cross-correlating sky emission lines between the target spectra and the
spectrum of target~5 {\it prior} to the sky subtraction data reduction
step. The typical size of this correction was $\pm 3$\kms, reflecting
inaccuracies in the wavelength calibration, possibly due to flexure in
the spectrograph during the 90 minutes of observation for each mask.
A heliocentric RV zeropoint was estimated by cross-correlating stars of
type M4 or later with archival VLT UVES spectra of the M4V--M6V
stars GL\,402, GL\,406 and GL\,876 for which precise heliocentric RVs
are known (see Bailer-Jones 2004 for details).

Heliocentric RVs versus spectral type are plotted in Fig.~\ref{rvspt}. 
Typical internal uncertainties are
of order a few \kms. Most objects are closely clustered in this
diagram. However, there is a clear upward trend towards later spectral
types that seems to be a consequence of spectral type mismatch between
target and template, probably due to our reliance on broad molecular
bands rather than sharp atomic lines in the later-type stars. Our best
estimate for the true heliocentric RV of the cluster is $+16.8\pm
1.1$\,\kms\ from the Li-rich stars of spectral type M4 or later. The
quoted error includes the scatter in our measurements and an estimate of the
external uncertainty judged from the variance of results using the
three different standard star templates. Our result agrees with (though is
much more precise than) the $+16.6\pm 6.0$\,\kms\ quoted by Rastorguev
et al. (1999).

\subsection{Cluster Membership}
\label{member}

\begin{figure}
\includegraphics[width=75mm]{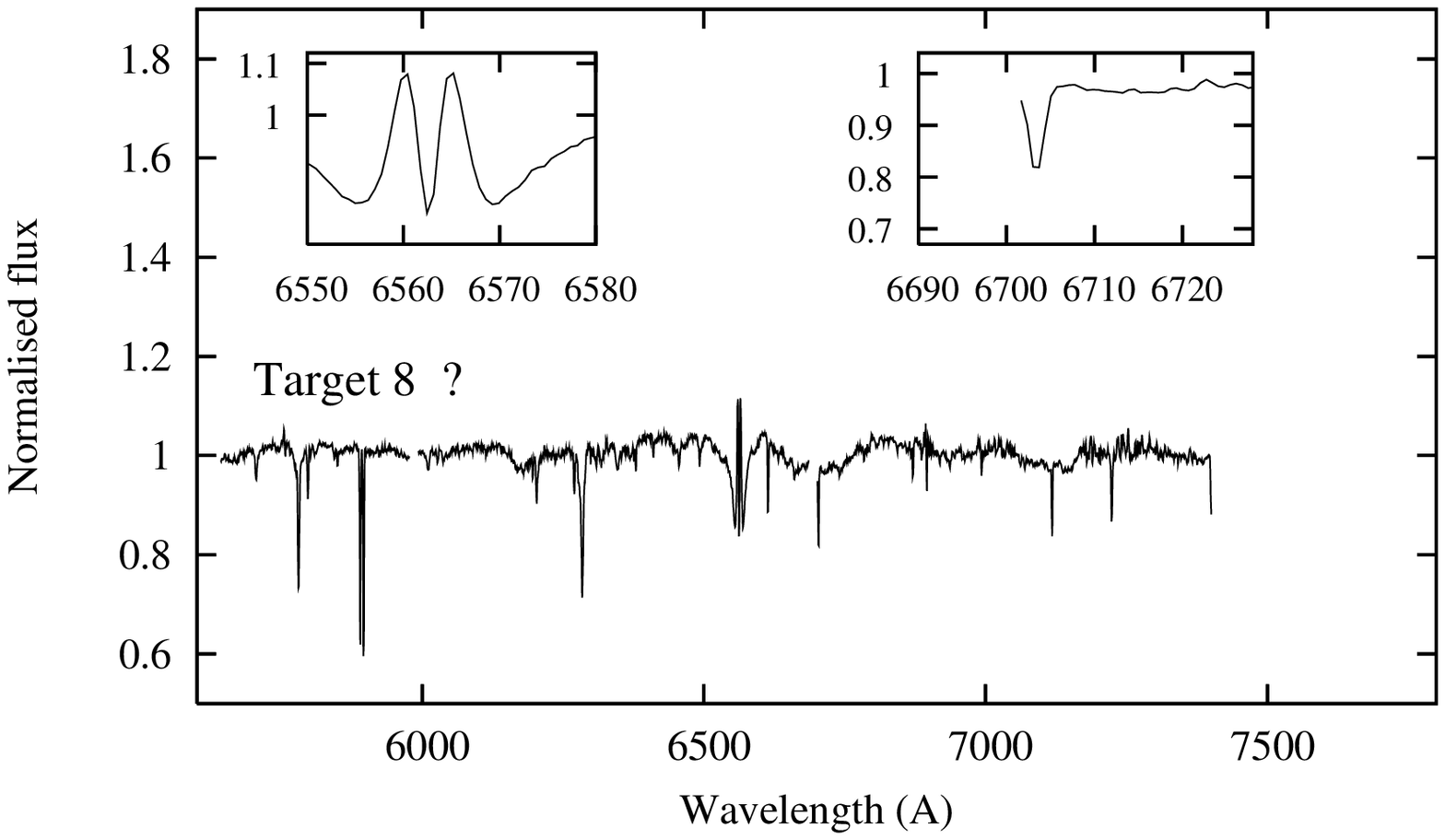}
\includegraphics[width=75mm]{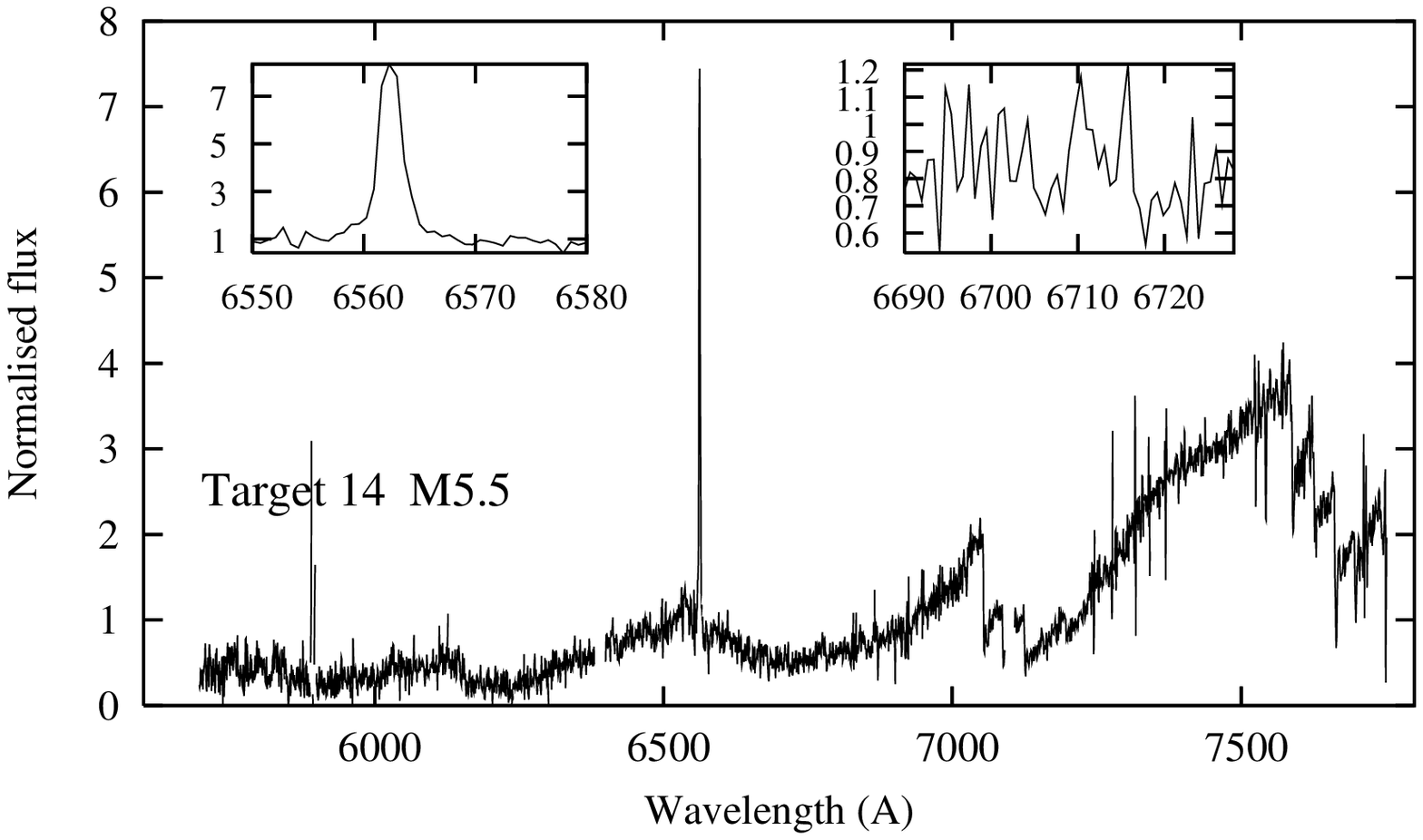}
\caption{Spectra for two targets with questionable cluster membership
  (see Section~\ref{member}). The inserts show normalised regions
  around the H$\alpha$ and Li\,{\sc i} features.
}
\label{strangespectra}
\end{figure}

The presence of a strong (EW$>0.3$\AA) \lii~6708\AA\ feature is taken
as a positive indicator of membership and is entirely supported by the
chromospheric H$\alpha$ emission exhibited by all these Li-rich
objects. The Li-rich candidates also have RVs with a very small
intrinsic dispersion (less than a few \kms\ after taking account of the
broad trend with spectral type) that is consistent with cluster
membership.

All three of the M dwarfs with no
Li measurement have H$\alpha$ emission lines consistent with the
Li-rich targets. However H$\alpha$ emission is no guarantee of extreme youth
among M dwarfs and one of these three objects has an RV inconsistent
with cluster membership, unless it is a short period binary system. For
now we regard all three objects as questionable members.

In order to investigate the possibility of anomalously large Li
depletion, we also consider those targets where Li was undetected.  Of
eight such targets, five have an H$\alpha$ line in absorption or barely
in emission. These are extremely unlikely to be young objects and all
have an RV which is inconsistent with cluster membership. These five
are all classified as definite non members. One object has no H$\alpha$
measurement and an RV inconsistent with cluster membership and we also
classify this as a definite non-member. The remaining two objects have
unusual spectra (see Fig.~\ref{strangespectra}) which are discussed
below.
 
Target 8. This star shows double peaked central H$\alpha$ emission
within a broad absorption feature. The spectrum is dominated by several
strong absorption lines at 5780.5\AA, 5796.9\AA, 6283.9\AA, 6613.7\AA\
and 7224.0\AA, which can be identified with diffuse interstellar
absorption bands (DIBS). There is no Li at 6708\AA\ although there is
an unidentified line at 6703\AA. The EWs of the DIBS imply quite a
high reddening. Using the relationships given by Jenniskens \& D\'esert
(1994), we estimate $E(B-V)\simeq 0.9$. From this, and also from the
position of the star in the $J-H$ versus $H-K$ diagram, we infer that
it has a spectral type of late A or early F and that the spectral type
of K5 assigned in Section~\ref{spt} must be in error. The star lies
blueward of the best fitting cluster isochrone in the $I_C$ versus $R_C-I_C$
CMD and the presence of additional reddening would increase this
discrepancy. However, given that the H$\alpha$ emission profile
strongly suggests circumstellar material and possibly an accretion
disc, we cannot be sure that the intrinsic optical colours would be
representative of the photosphere of a late A star in any case. Hence
this object could be a young Herbig Ae star either within NGC~2169 or
more likely at a much greater distance. Given this uncertainty, we will not
consider the star as a cluster member in what follows.

Target 14. This object has an upper limit to its Li EW which could just
 be consistent with the presence of significant Li in the
 photosphere. Indeed, there is a hint of an Li line at the 1-2 sigma
 level. Its EW[H$\alpha$] is the largest in our sample and close to the
 empirical border between CTTS and WTTS. Its RV is consistent with
 cluster membership. The width of the H$\alpha$ line is only $\simeq
 180$\kms\ at 10 per cent of maximum. The status of this star is
 questionable. It is either a cluster member or a very active
 foreground dMe star. A better spectrum of the \lii~6708\AA\ line is
 needed and it is not considered as a cluster member in the analysis
 that follows.

\section{Age estimates}
\label{age}

Absolute ages for open clusters are usually model dependent.  This
is especially true in very young open clusters and star forming
regions. A valuable exercise is to compare cluster ages derived from
techniques that rely on different aspects of stellar physics. Agreement
would instil confidence in the accuracy of stellar ages whilst
discrepancies would highlight potential weaknesses in our understanding
of stellar evolution.

Previous work has taken ages determined by fitting model isochrones
to the positions of high-mass stars that have undergone nuclear
evolution in the Hertzsprung-Russell (H-R)
diagram, and compared them 
with ages determined from isochrones which trace the descent
of low-mass PMS stars contracting towards the hydrogen burning main
sequence. In clusters with ages of 50--700\,Myr good agreement has been
claimed (e.g. Lyra et al. 2006), but in younger clusters
($<30$\,Myr) discrepancies are more common (e.g. Piskunov et al. 2004).
Here, ages from evolved high-mass stars are often more uncertain
because there are fewer such objects and the ages are dependent on
details such as rotation and the degree of mixing in the convective
core (Chiosi, Bertelli \& Bressan 1992; Meynet \& Maeder 1997,
2000). The ages obtained from fitting low-mass isochrones also become
more model dependent -- the details of the stellar atmospheres,
interior convection and even the initial conditions become increasingly
important (see Baraffe et al. 2002).

A third technique has begun to be added to these comparisons. The
abundance of Li in convective envelopes and atmospheres is sensitively
dependent to the temperature at the base of the convection zone (or
stellar centre in the case of fully convective PMS stars). Once at a
threshold temperature of about $3\times10^{6}$\,K, Li is burned rapidly
in (p,$\alpha$) reactions. The mass and hence luminosity and
temperature at which Li burning commences is age dependent and
isochrones of Li depletion can be used to estimate the ages of clusters
(see Jeffries 2006 for a review).  This technique has been used to
obtain precise ages for several open clusters. Isochronal ages from
high-mass stars agree with the lithium depletion ages for open clusters
in the range 50--150\,Myr providing that some extra mixing (caused by
core overshoot or rotation?) extends the main sequence lifetimes of
5--8\,$M_{\odot}$ stars (Stauffer, Schultz \& Kirkpatrick 1998;
Stauffer et al. 1999; Barrado y Navascu\'es, Stauffer \& Jayawardhana
2004). Jeffries \& Oliveira (2005) have also shown that the lithium
depletion age of a 35\,Myr cluster agrees with an age derived from
isochronal fits to low-mass PMS stars in the same cluster.  However,
discrepancies have also been reported. Song, Bessell \& Zuckerman
(2002) and White \& Hillenbrand (2005) have reported on individual
PMS stars which appear to show much more Li depletion than expected for
their isochronal ages in the H-R diagram.

\subsection{Isochrone matches to high mass stars}
\label{himassage}

\begin{figure*}
\centering
\begin{minipage}[t]{0.45\textwidth}
\includegraphics[width=71mm]{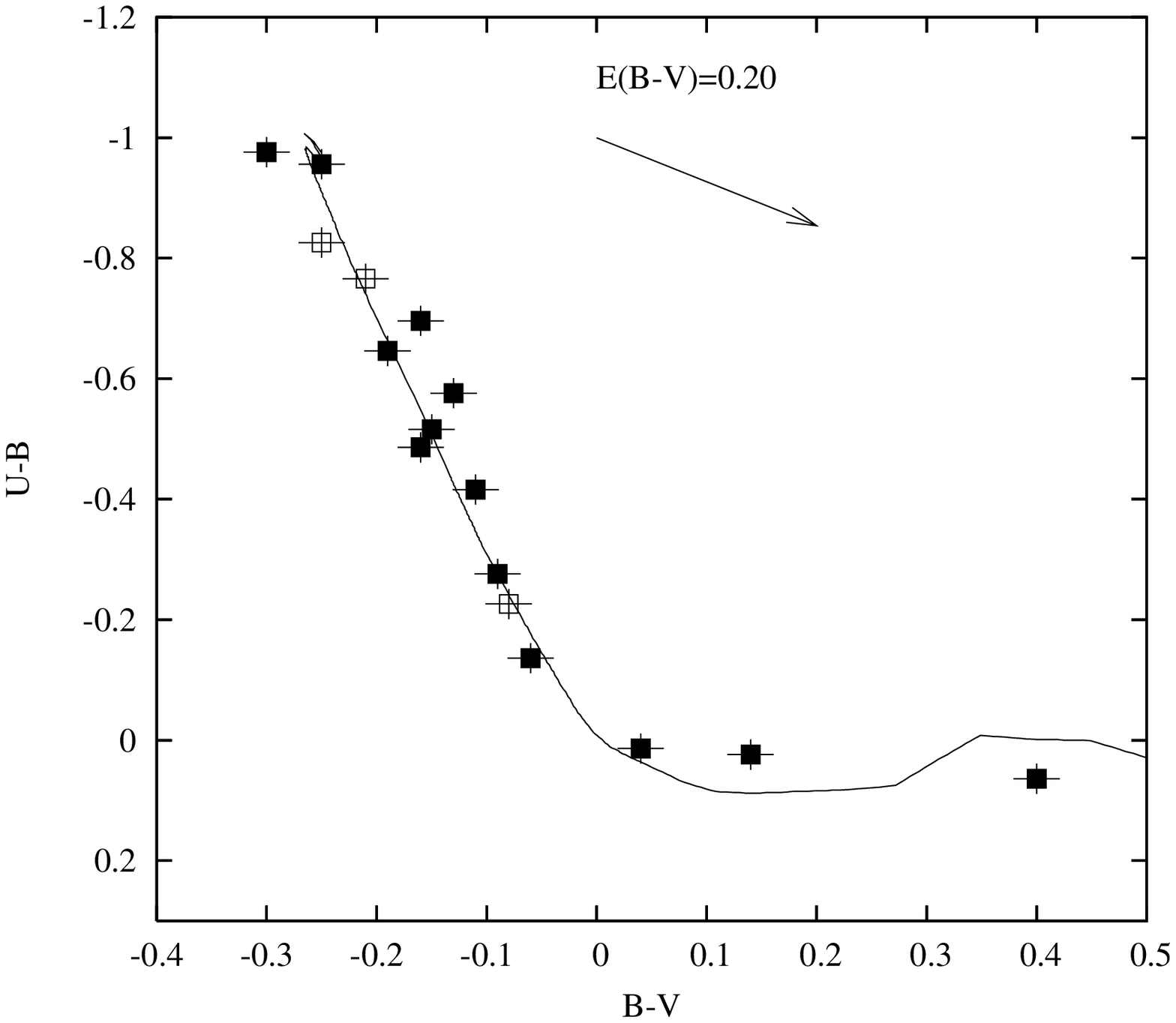}
\end{minipage}
\begin{minipage}[t]{0.45\textwidth}
\includegraphics[width=71mm]{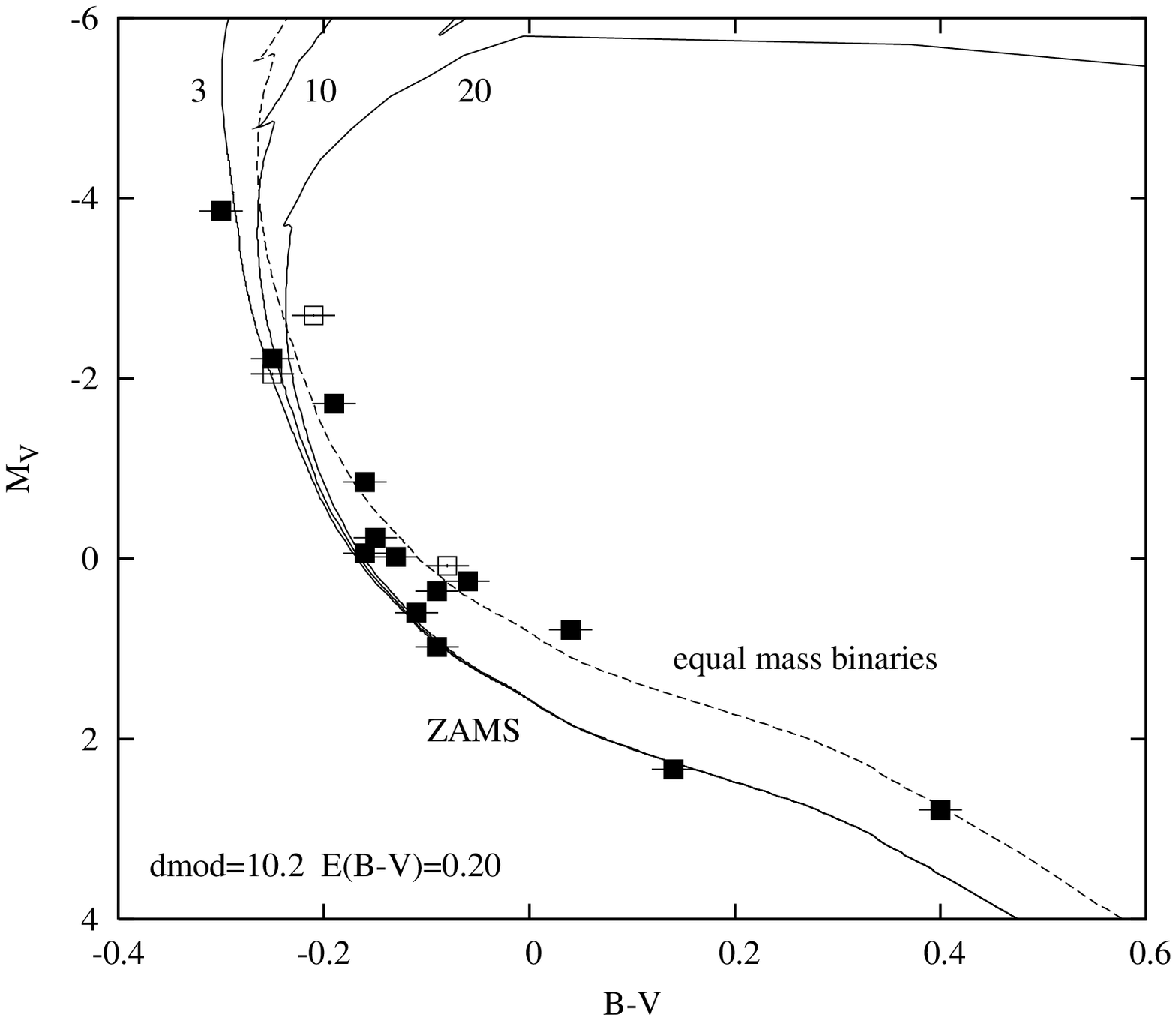}
\end{minipage}
\includegraphics[width=71mm]{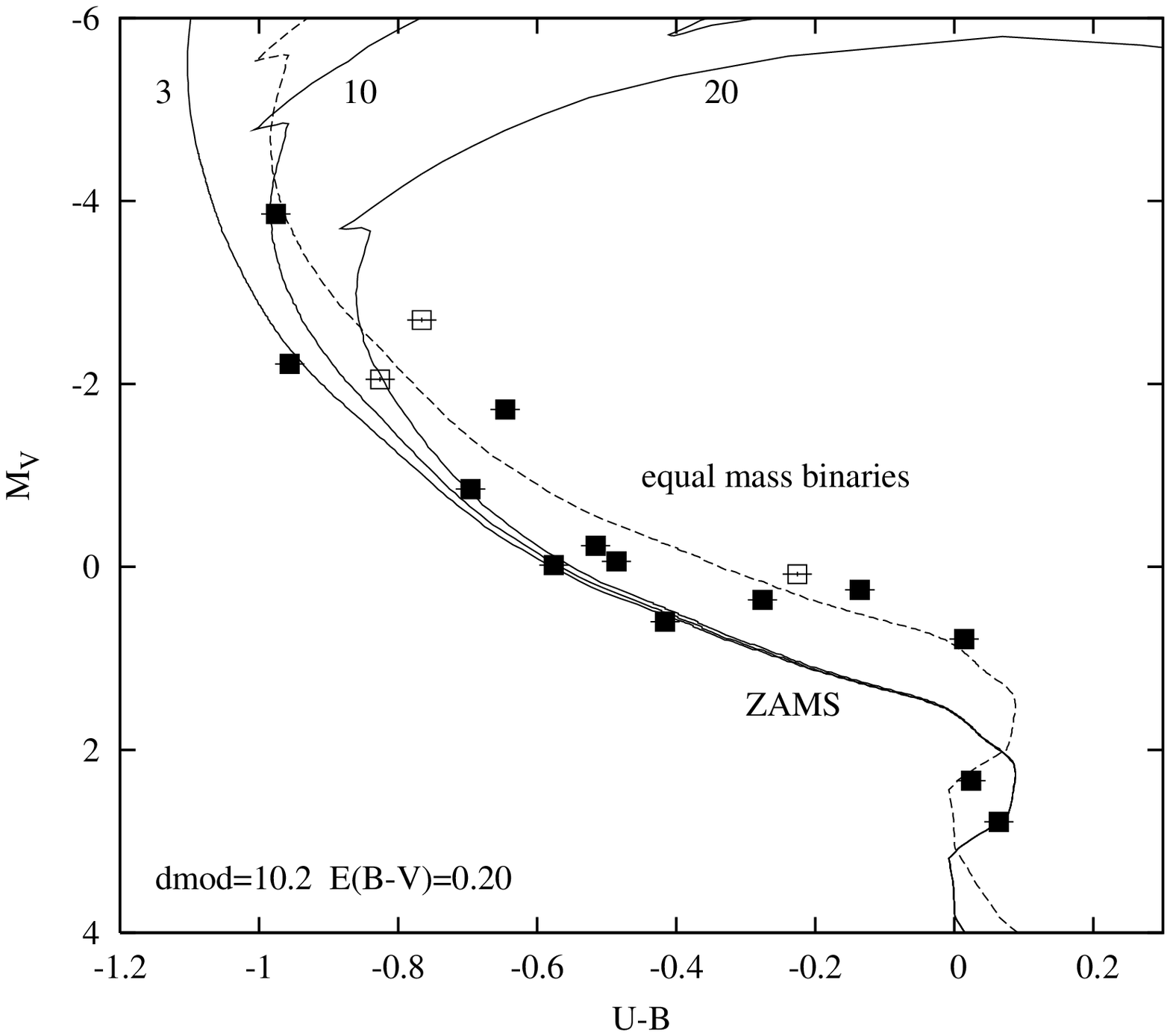}
\caption{The intrinsic colour-colour and colour-magnitude diagrams for early-type
  members of NGC 2169. Data from Sagar (1976) are plotted. Open symbols
  indicate objects whose membership status is uncertain or
  disputed. Geneva model isochrones are from Lejeune \& Schaerer (2001).
  (a) A 10\,Myr isochrone is plotted after a reddening of
  $E(B-V)=0.20$ and $E(U-B)=0.146$ has been subtracted from the
  data. (b and c) Isochrones
  with ages of 3, 10 and 20\,Myr are
  shown. The additional dashed line indicates the equal-mass binary sequence for
  the 10\,Myr isochrone.
}
\label{ubvplot}
\end{figure*}

The high-mass population of NGC~2169 is relatively sparse with only one
clearly evolved B2\,III binary star (Abt 1977). Age constraints from
these stars will therefore be quite poor, but we can still get valuable
constraints on the cluster distance that will be useful when
considering low-mass isochronal fits.  Perry et al. (1978) based their
age on what they consider to be the brightest, unevolved, non-binary
member of NGC 2169 -- namely Hoag 6, with a spectral type of B3V (Abt
1977). A calibration of the duration of main sequence burning and
absolute magnitude due to Schlesinger (1972) was used, leading to an
upper age limit of 23\,Myr. Delgado et al. (1992) claim a similar status for
Hoag 4 with a spectral type of B2.5IV (Abt 1977) or B2V (Harris 1976)
and hence determine a slightly lower upper age limit of 16\,Myr.

We have re-evaluated the reddening, distance and age of NGC 2169 using
the $UBV$ data set published by Sagar (1976) together with the
solar metallicity evolutionary models of Schaller et al. (1992) which
have been transformed into isochrones in the Johnson $UBV$ system by Lejeune \&
Schaerer (2001 -- henceforth known as the Geneva models and isochrones).
The Sagar (1976) dataset
is the largest homogeneous set of photoelectric data for NGC
2169. We reviewed the cluster membership using information in Sagar
(1976), Perry et al. (1978) and Delgado et al. (1992). There was
generally consensus over membership, but in three cases (Hoag 2, 5, and
15) membership is disputed.

The intrinsic $U-B$ vs $B-V$ colour-colour diagram, $V$ vs $B-V$ and
$V$ vs $U-B$ CMDs are shown in Fig.~\ref{ubvplot}. We find that a
reddening of $E(B-V)=0.20\pm 0.01$ applied to the 10\,Myr Geneva
isochrone satisfactorily models the colour-colour diagram. Note that we
only consider the age-independent main sequence portion of the $U-B$ vs
$B-V$ diagram below the ``blue turn-off'' and exclude the bluest
(evolved) star from the fit. As this portion of the colour-colour locus
is age independent for ages $\leq 20$\,Myr, then the reddening estimate
is also independent of assumed age in this range (see below).  Then,
assuming that $A_V/E(B-V)$=3.10, the $V$ vs $B-V$ CMD can be matched
(by eye) with an intrinsic distance modulus of $10.2\pm0.2$. From this
CMD the age of the cluster is certainly less than 20\,Myr based on the
two brightest undisputed members and could be less than 10\,Myr based
only on the brightest object.  The brightest star in the cluster is
actually a close-to-equal mass binary system but this does not affect
the age limit because the 10\,Myr isochrone is almost vertical at this
magnitude. In principle $V$ vs $U-B$ could offer better distance
precision because the ZAMS locus has a shallower gradient for B stars
in this CMD. However, we find the photometry appears more scattered,
with points lying well outside a plausible band that could be explained
by equal mass binary systems.  In this CMD a less certain distance
modulus of $10.2\pm 0.3$ and an age of $\simeq 10$\,Myr seem
appropriate.  An age of $\geq 20$\,Myr or more is still ruled out by
the brightest pair of undisputed members.

\label{ms}

\begin{figure}
\includegraphics[width=75mm]{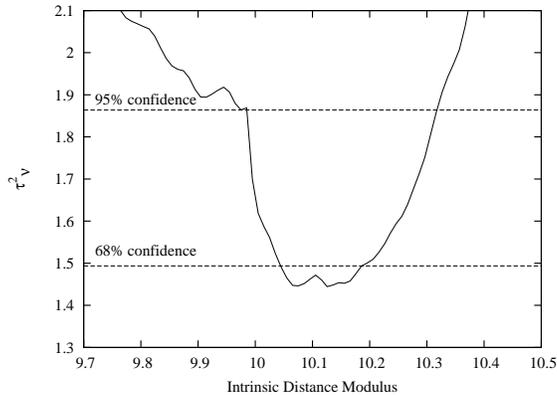}
\caption{
$\tau_{\nu}^2$ as a function of distance for the fit to the 10 Myr Geneva
isochrone.
The 68 and 95 per cent confidence levels are marked.
}
\label{GENEVA:grid}
\end{figure}

\begin{figure}
\includegraphics[width=75mm]{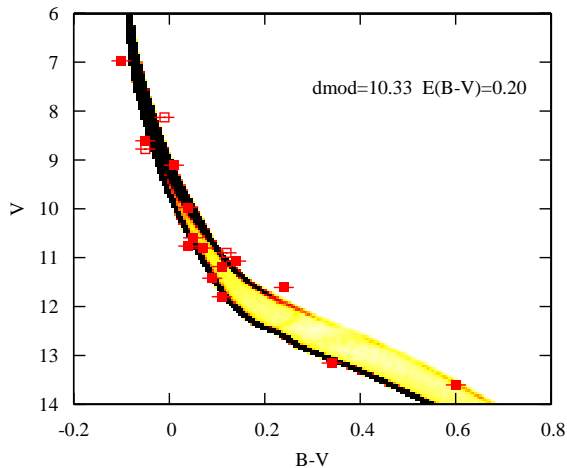}
\caption{The best fitting Geneva model with a 50 per cent binary
frequency is represented with a greyscale distribution.
The data points from Sagar (1976) are shown with symbols as in
Fig.~\ref{ubvplot}. The brightest cluster member was not fitted (see text).
}
\label{GENEVA:best_small}
\end{figure}

To put the distance estimate on a firmer basis we fitted the $V$ versus
$B-V$ data using the $\tau^2$ technique described in detail by Naylor
\& Jeffries (2006). $\tau^2$ is a generalisation of $\chi^2$ that can be
used to fit data points to a two-dimensional distribution, hence
allowing for the binary content of a cluster sequence in the CMD, and
provides robust, statistically meaningful error bars.  We assumed that
a 10\,Myr Geneva isochrone was appropriate and a binary fraction of 50
percent, with secondary stars in the binaries randomly selected from a
uniform mass distribution between the mass of the primary and zero.  As
discussed in Naylor \& Jeffries (2006), the precise value of the binary
fraction has only a small impact on the derived parameters. In this
case the intrinsic distance modulus of the cluster was the only free
parameter and we fitted to all data points apart from the brightest,
Hoag 1, in order to ensure that the derived distance was insensitive to
the assumed age\footnote{In fact we attempted a search for the
best-fitting distance {\em and} age simultaneously from all the
high-mass stars (as in Section~\ref{lowmassage}), but still found that
the distance estimate was almost independent of age between 0 and
20\,Myr.}. We searched in distance modulus, to yield the $\tau^2$ plot
shown in Figure \ref{GENEVA:grid} and a best-fit distance modulus of
10.13 (in reasonable agreement with the ``by-eye'' result above), with
a 68 percent confidence range of 10.04-10.19 (see Table \ref{fits}).
The data and best-fitting model are shown in Figure
\ref{GENEVA:best_small}.  We obtained a $P_r(\tau^2)$ (equivalent to
the probability of exceeding a given $\chi^2$) of 0.03, indicating that
the model just about provides a satisfactory representation of the data
and that the uncertainties are approximately correct. Adding a
systematic error of 0.02 mag to each data point would bring
$P_r(\tau^2)$ up to 0.5, increasing the distance modulus uncertainty
only slightly.

\subsection{Low mass isochrones}
\label{lowmassage}

We can also estimate an age by fitting the positions of the
spectroscopically confirmed low-mass members of NGC~2169 with
isochrones from various PMS evolutionary models.  The age
derived in this way is {\it strongly} correlated with the assumed distance to
the cluster.  To make the correlation explicit we again carried out
$\tau^2$ fits, this time allowing both age and distance modulus to be free
parameters.  We can then use this relationship along with the distance
derived from the high mass stars to find a best estimate for the age of
NGC~2169.  Significantly
different ages (in the formal statistical sense) are found depending on which
model we use.  In what follows we therefore consider each model in
turn.  For each model isochrone, effective temperature and bolometric
luminosity were converted into $I_C$ and $R_C-I_C$ using
empirical relationships between colour and temperature and colour and
bolometric correction. The colour-temperature relationship was
established for each set of evolutionary models by demanding that an
isochrone with an age of 120\,Myr at a distance modulus of 5.6 and with
an extinction $E$($R_C-I_C$)=0.029 matches the $I_C$ versus $R_C-I_C$
locus of the Pleiades (see Jeffries, Thurston \& Hambly 2001; Naylor et
al. 2002; Jeffries \& Oliveira 2005).  For NGC~2169 we used an
extinction of $E$($R_C-I_C$)=0.14, and for both clusters assumed
$A_I/E(R_C-I_C)$=2.57 (Dean et al. 1978; Taylor 1986).  The bolometric
correction was assumed to be a function of $R_C-I_C$, and obtained by
fitting the colours of late-type dwarfs, as in Naylor et al. (2002).
The main assumption in this method of ``tuning'' the isochrones is that
the relationships between temperature, bolometric correction and colour
are the same at 120\,Myr and at the ages of the isochrones that fit
NGC~2169 (see Jeffries \& Oliveira 2005 for a discussion of this point).

Since we are using PMS contraction to derive the age of
the cluster, we wish to fit the lowest-mass objects possible.  The best
dataset to fit therefore is our own $I_C$ vs $R_C-I_C$ photometry of
members confirmed in Section~\ref{member}.  We fitted the 36 stars we
identified as members, with the exception of target 45, which is marked
as non-stellar in our catalogue due to the presence of a nearby bright
star.

\subsubsection{The Baraffe isochrones}

\begin{figure*}
\centering
\begin{minipage}[t]{0.45\textwidth}
\includegraphics[width=71mm]{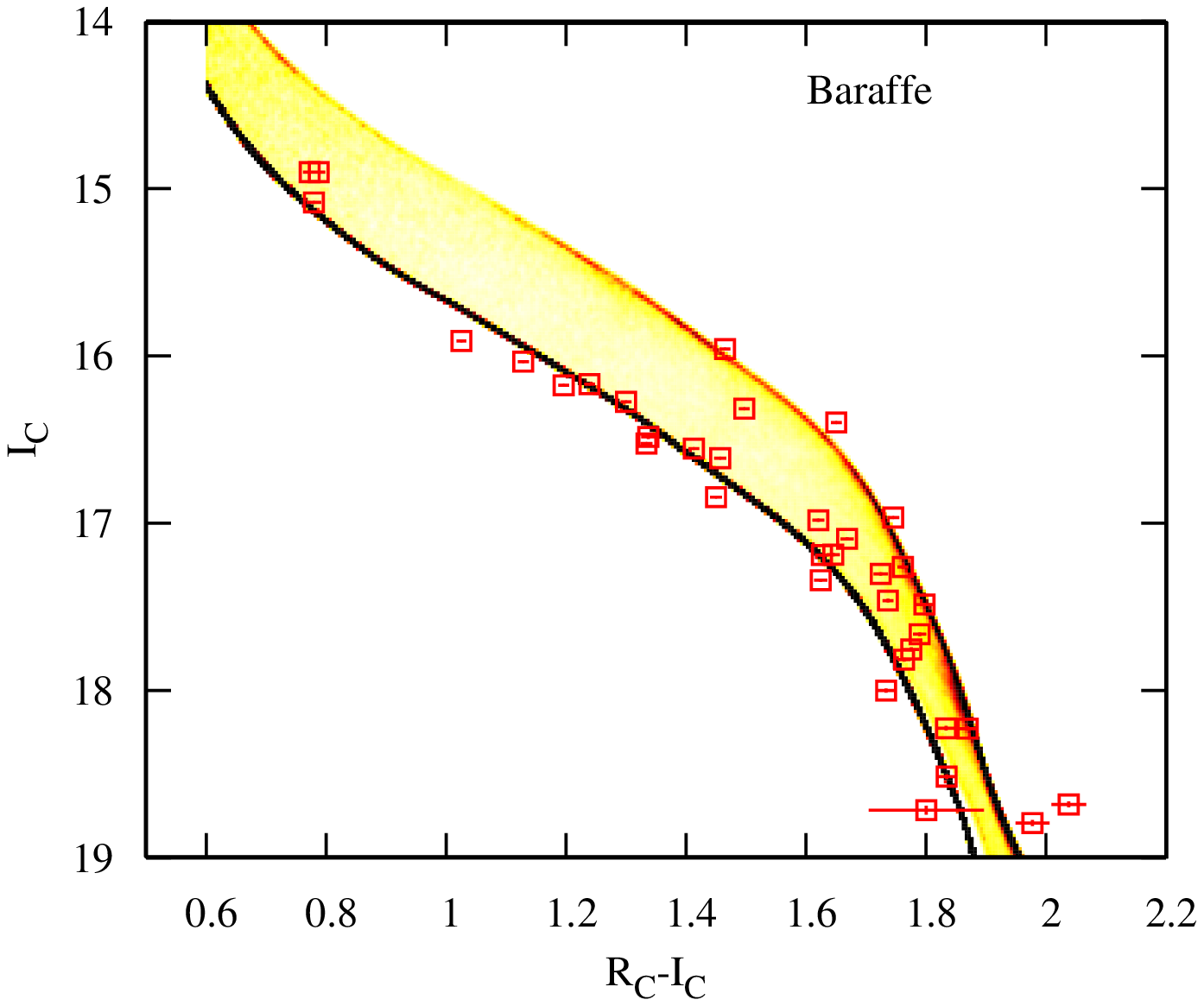}
\includegraphics[width=71mm]{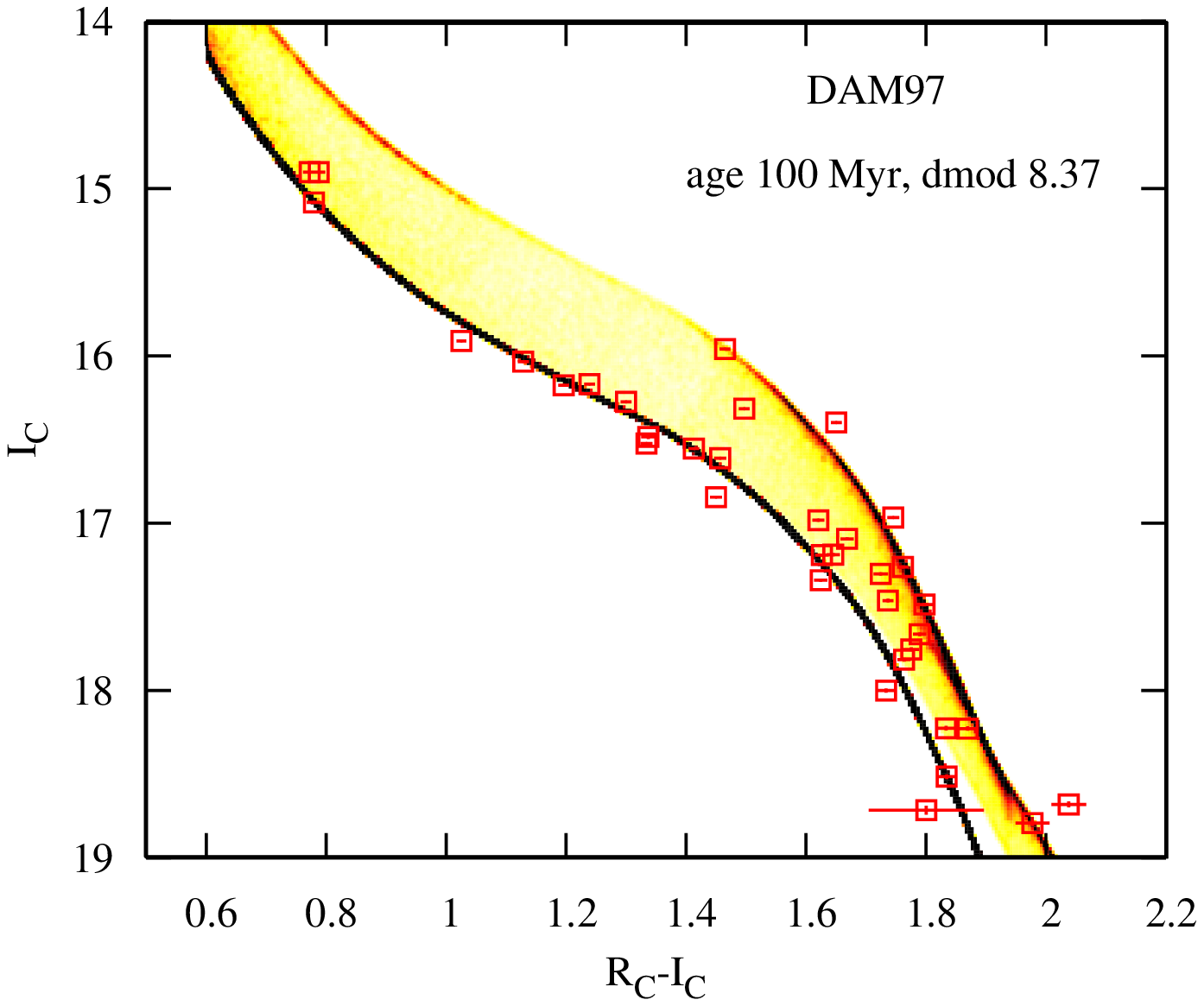}
\end{minipage}
\begin{minipage}[t]{0.45\textwidth}
\includegraphics[width=71mm]{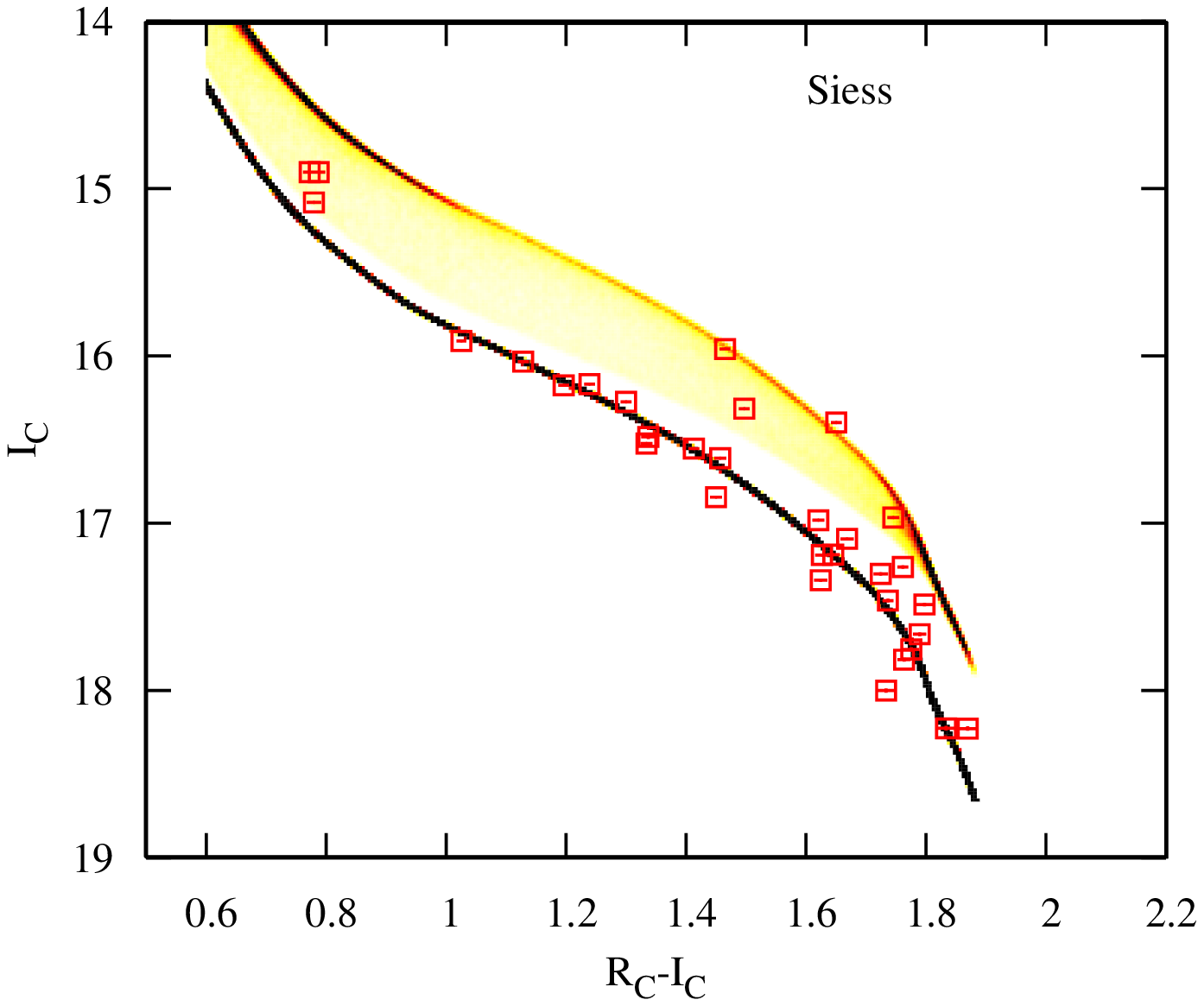}
\includegraphics[width=71mm]{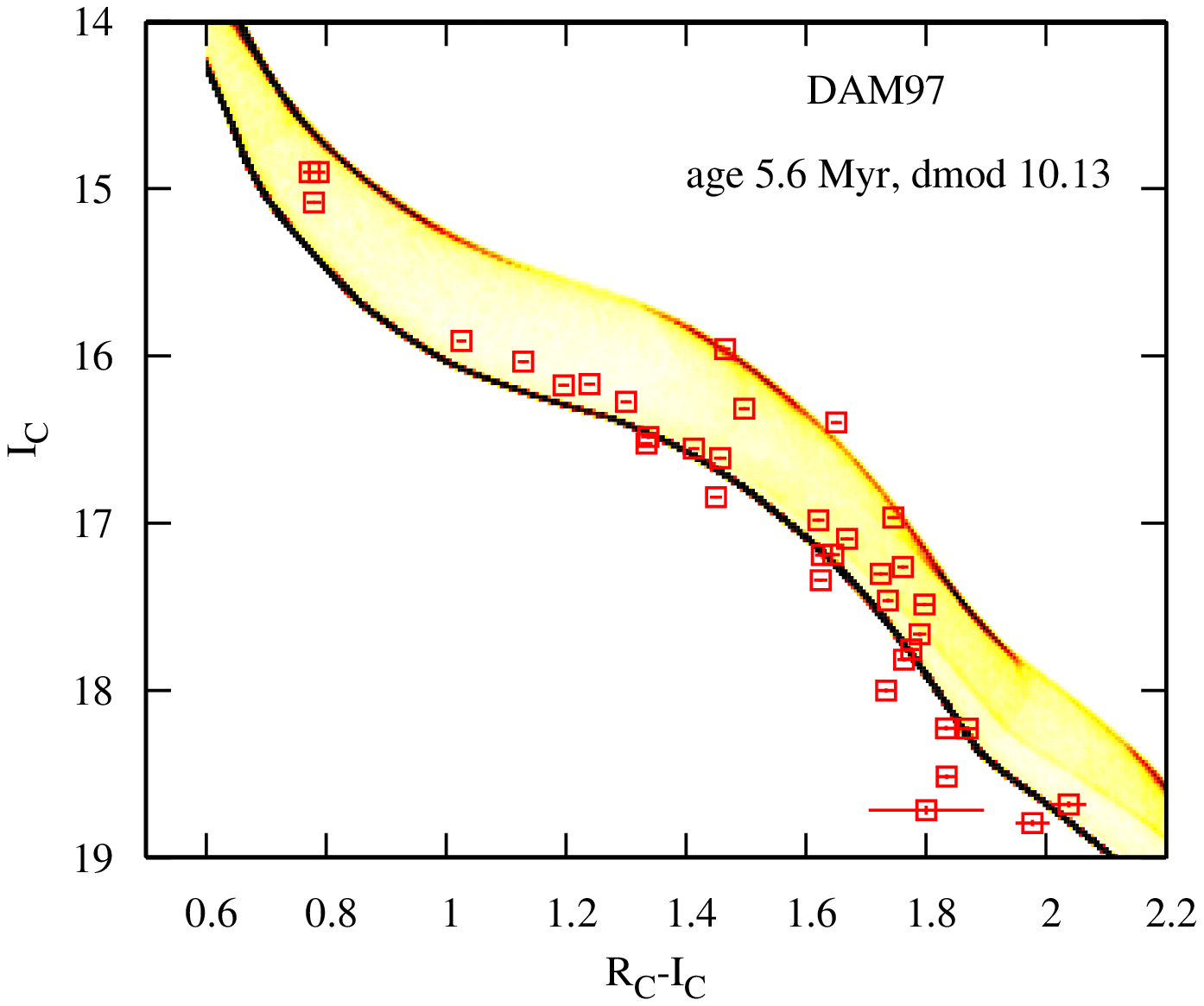}
\end{minipage}
\caption{The best fitting distributions using isochrones from
several evolutionary models, a binary frequency of 50 per cent and a
flat mass ratio distribution. The fitted data points are shown with
error bars that {\it do not} include the 0.03 mag systematic that is
discussed in the text. The best-fitting age and distance are given in
Table~\ref{fits}. (a) Isochrone from Baraffe et al. (1998, 2002)
using a mixing length of 1.0 pressure scale heights; (b) Isochrone
from Siess et al. (2000) models with a metallicity of
0.02 and no convective overshoot; (c) Isochrone from D'Antona \&
Mazzitelli (1997); (d) Isochrone from D'Antona \& Mazzitelli (1997),
but in this case the distance modulus is constrained to be 10.13 mag,
the best-fit value from the high-mass main sequence (see
section~\ref{himassage}).}
\label{bestfits}
\end{figure*}

Our baseline models were the isochrones of Baraffe et al. (1998, 2002) (using
the solar metallicity models with a convective 
mixing length set to 1.0 pressure scale heights),
with an assumed binary fraction of 50 percent and uniform mass ratio
distribution. Studies of the field binary population suggests these are
reasonable assumptions for low-mass stars (e.g. Duquennoy \& Mayor
1991; Fischer \& Marcy 1992).
Secondary stars which lie below the lowest mass available in 
the models were assumed to make a negligible contribution to the system 
light, which is equivalent to placing the binary on the single-star sequence.
This limitation of the isochrones leads to the wedge of zero probability 
between the single and binary star sequences at low masses visible in Figure
\ref{bestfits}.

\begin{figure*}
\centering
\begin{minipage}[t]{0.32\textwidth}
\includegraphics[width=55mm]{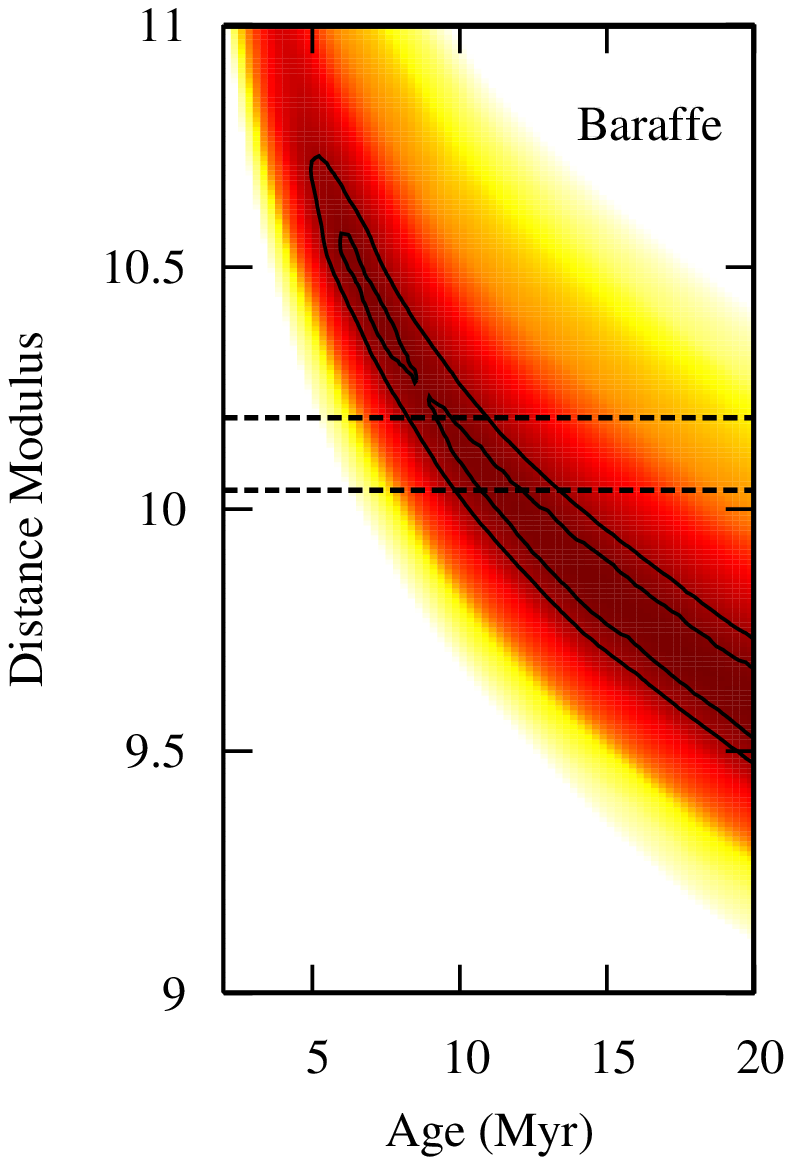}
\end{minipage}
\begin{minipage}[t]{0.32\textwidth}
\includegraphics[width=55mm]{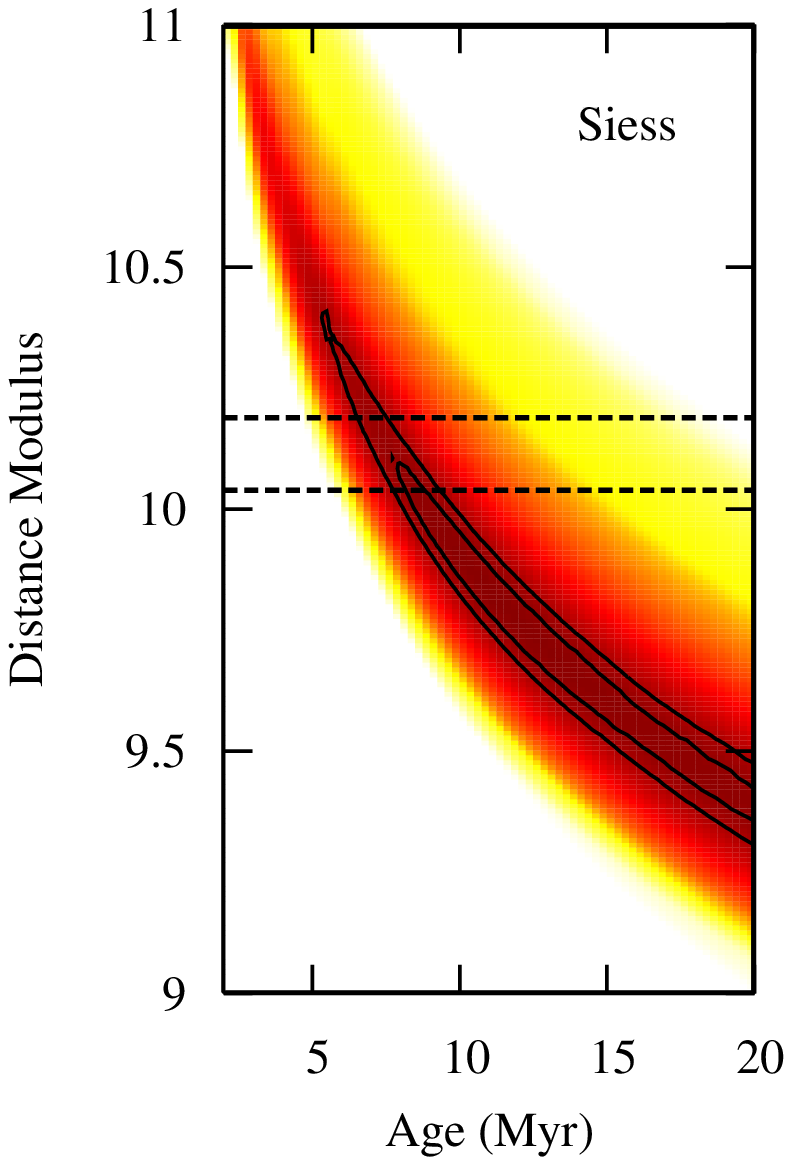}
\end{minipage}
\begin{minipage}[t]{0.32\textwidth}
\includegraphics[width=55mm]{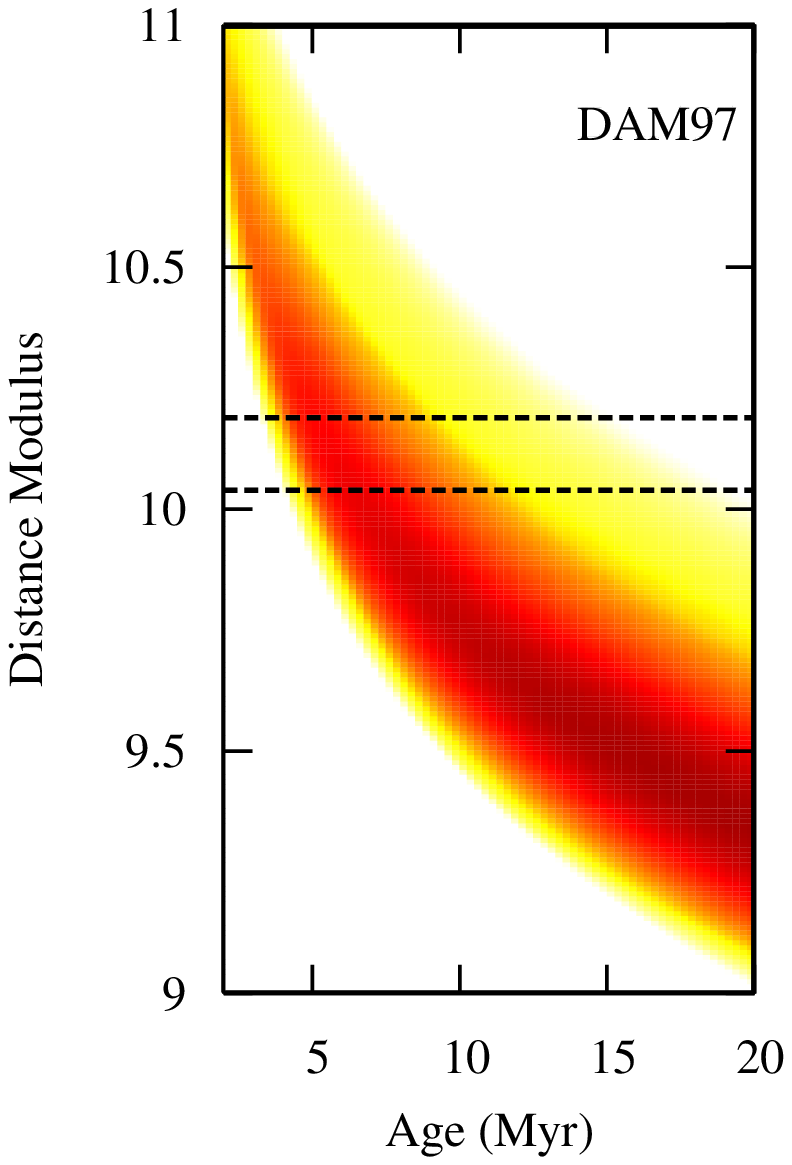}
\end{minipage}
\caption{The $\tau^2$ grid for the low-mass isochrone fits.
The contour lines are the 68 and 95 percent confidence limits.
The horizontal lines are the 68 percent confidence limits for
the distance modulus derived from the main-sequence fitting to the
higher mass stars (see section~\ref{himassage}.
(a) Baraffe et al. (1998, 2002) models with a mixing length of 1.0 pressure
scale heights. (b) The Siess et al. (2000) models with a metallicity of
0.02 and no convective overshoot. (c) The D'Antona \& Mazzitelli (1997)
models. In this latter case the confidence limits lie at larger ages than displayed
on the grid (see text).
}
\label{tauplot}
\end{figure*}

A straightforward fit of this type leads to a $P_r(\tau^2)$ of 
0.0025, after clipping out two data points.
This is unacceptably low, and already includes the addition of a
magnitude-independent uncertainty
(0.01 mag in $R_C-I_C)$ to all the data 
points to allow for the uncertainties in our profile correction and 
transformation to the standard system (see section \ref{ccdphotom}).

To account for this extra scatter and hence derive reliable
uncertainties on the derived parameters, we increased the systematic
uncertainties in the $R$ and $I_C$ measurements to 0.03 mag and obtain
an acceptable $P_r(\tau^2)=0.47$.  This procedure can be considered
analogous to increasing the uncertainties on each point in order to
obtain $\chi^2_{\nu} \simeq 1$.  The resulting parameters are given in
Table \ref{fits} and the best fit is shown in Figure
\ref{bestfits}a.  The fit is a reasonable one, with the
model explaining most of the spread in colour as due to binarity. Only
at the faintest magnitudes is there a hint of any problem.

The most important result, however, is the $\tau^2$ grid of Figure 
\ref{tauplot}a, since we can combine it with the
age constraint from the high-mass stars (Section \ref{ms}).
If we choose the lowest values of $\tau^2$ at the 68 percent
confidence-limit distances, we obtain the age constraints given
in Table \ref{fits}.

\begin{table}
\caption{
The parameters derived from pre-main-sequence fits.
}
\begin{tabular}{ccccccccccccccccc}
\hline
        & \multicolumn{2}{c}{Best Fit} &  \multicolumn{2}{c}{Age (Myr) for $d_m$} \cr
Model   &  $d_m$ (mags) & Age (Myr)    &  10.19  & 10.04  \cr
\hline
Geneva  &  10.13      &\cr
        &\multicolumn{2}{c}{(10.04-10.19 68\%)}\cr
Baraffe & 9.5  & 24   & 9.5 & 11.5  \cr
Siess   & 9.8  & 11   & 7.0 &  8.5  \cr
DAM97   &      & $>$100 & 5.0 &  6.5  \cr
\hline
\end{tabular}
\label{fits}
\end{table}

\subsubsection{The Siess isochrones}

We also fitted the data to the Siess et al. (2000) isochrones with
metallicty $Z=0.02$ and no convective overshoot.  
The resulting fit and $\tau^2$ space are shown in
Figures \ref{bestfits}b and \ref{tauplot}b.  Here there is a
problem with the faintest objects, because the models go no lower than
0.1\,$M_\odot$.  This means that the ``binary wedge'' is very large,
with the equal-mass binary sequence disappearing completely at quite
bright magnitudes.  We therefore
removed the four objects fainter than $I_C$=18.3.  
We obtain a best fit at an age of 11\,Myr
and a distance modulus of 9.8. The  $P_r(\tau^2)$ is 0.87, 
which although nominally better than the value obtained for
the Baraffe et al. (1998, 2002) isochrones, one should recall excludes
a contribution from the faintest datapoints that contributed most to $\tau^{2}$
for the Baraffe models.

\subsubsection{The DAM97 isochrones}

Finally we carried out fits to the models of 
D'Antona \& Mazzitelli (1997 -- hereafter the DAM97 isochrones).
The formal fits to these isochrones imply NCG~2169 is very old
($>$100\,Myr).
This behaviour appears to be caused by the way the ischrones do not steepen
at $R_C-I_C > 1.7$ in contrast to the Baraffe and Siess isochrones.
Older isochrones tend to be straighter, which clearly lead to better
fits if the distance is unconstrained.
The $\tau^2$ grid for ages $\leq 20$\,Myr is shown in Figure
\ref{tauplot}c.
Although the best-fitting distance is clearly far removed from the
values given by the other models, the correct question to ask is
whether the distances are co-incident at a given confidence level.
As we have not found the lowest point in the $\tau^2$ space our normal
method of finding the contour will over-estimate its size.
Even then we find that the youngest age enclosed by the 95 percent 
confidence contour is 20\,Myr (at a distance modulus of 9.4).
The best fitting model for a
fixed distance modulus of 10.13 mag is shown in
Fig.~\ref{bestfits}d. As implied by Fig.~\ref{tauplot}c the fit is
much worse. Hence the DAM97 models are a poor
description of the combined dataset.

\subsubsection{The age of NGC~2169}

The numerical results of our fitting are summarised in Table
\ref{fits}. From this it can be seen that if the distance is
constrained to be that implied by fits to the high-mass stars, then 
the DAM97 isochrones yield
the youngest age, then Siess et al. (2000), and finally Baraffe et
al. (1998, 2002).
The ages derived using the 68 percent confidence interval distance
from the main-sequence fitting just fail to
overlap, and hence the uncertainty in age is dominated
by the choice of model, not by the data. 
Both the Baraffe et al. (1998, 2002) and Siess et al. (2000) models
satisfactorily fit the shape of the low-mass pre-main-sequence at the
distance implied by fits to the upper main-sequence. Thus our
best estimate for the age of NGC~2169 is $9\pm 2$\,Myr, the upper limit
defined by the Baraffe et al. models and the lower limit by the Siess et al. models.

That the DAM97 isochrones struggle to reproduce the low-mass pre-main
sequence unless the distance modulus is allowed to become unreasonably
low and the age unreasonably high does not necessarily rule them
out. The discrepancies between data and model occur predominantly in
the coolest stars where it is still possible that systematic
uncertainties in the photometric calibration (in this study or in the
Pleiades data which calibrates the colour-temperature relationship)
could change the data--model comparison. However, if the coolest stars
were to be made even redder to better match the isochrone shapes then
the derived age for the cluster would be even younger than 5\,Myr.

\subsection{The Li depletion age}

\label{limassage}
\begin{figure*}
\centering
\begin{minipage}[t]{0.45\textwidth}
\includegraphics[width=71mm]{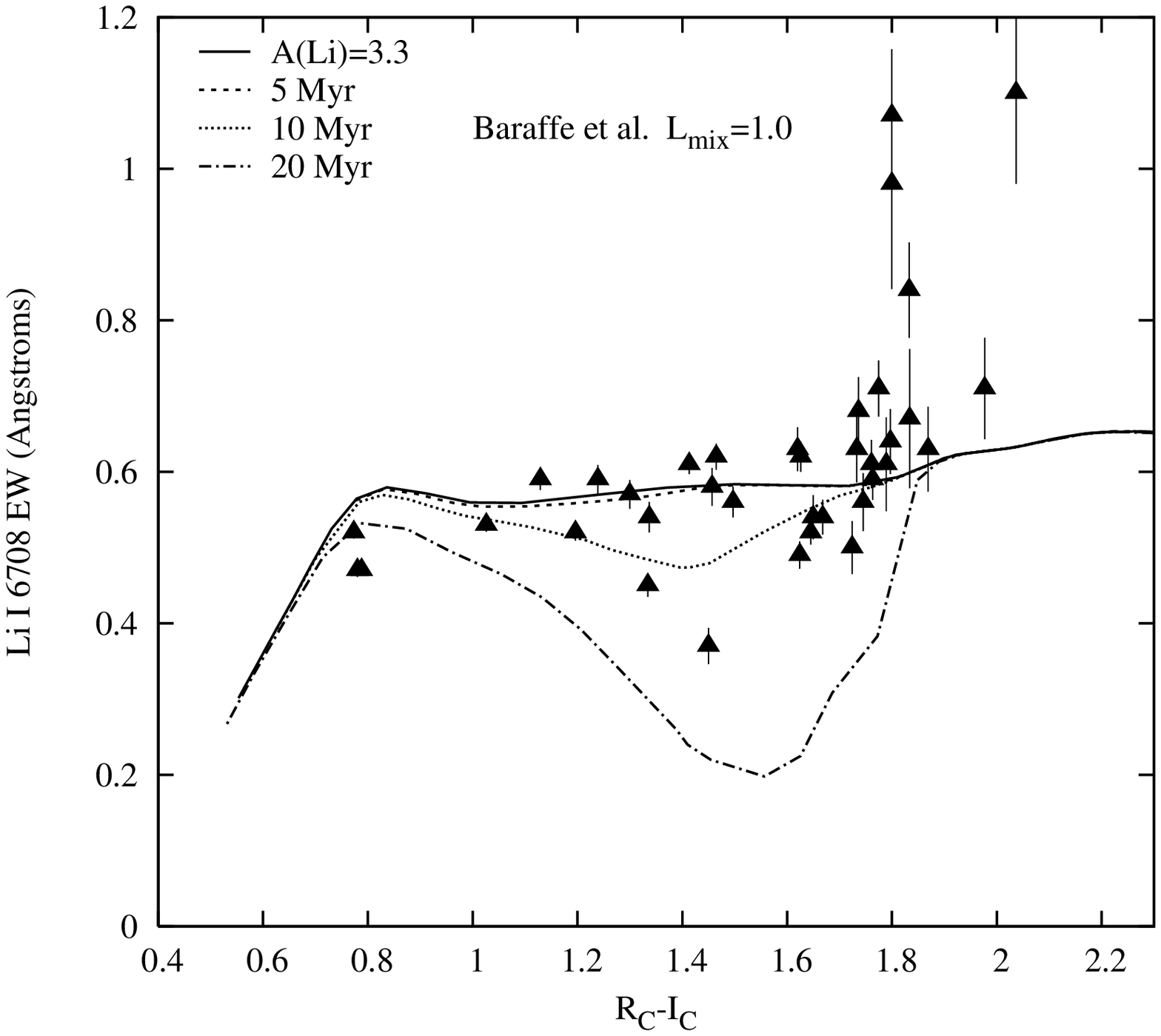}
\end{minipage}
\begin{minipage}[t]{0.45\textwidth}
\includegraphics[width=71mm]{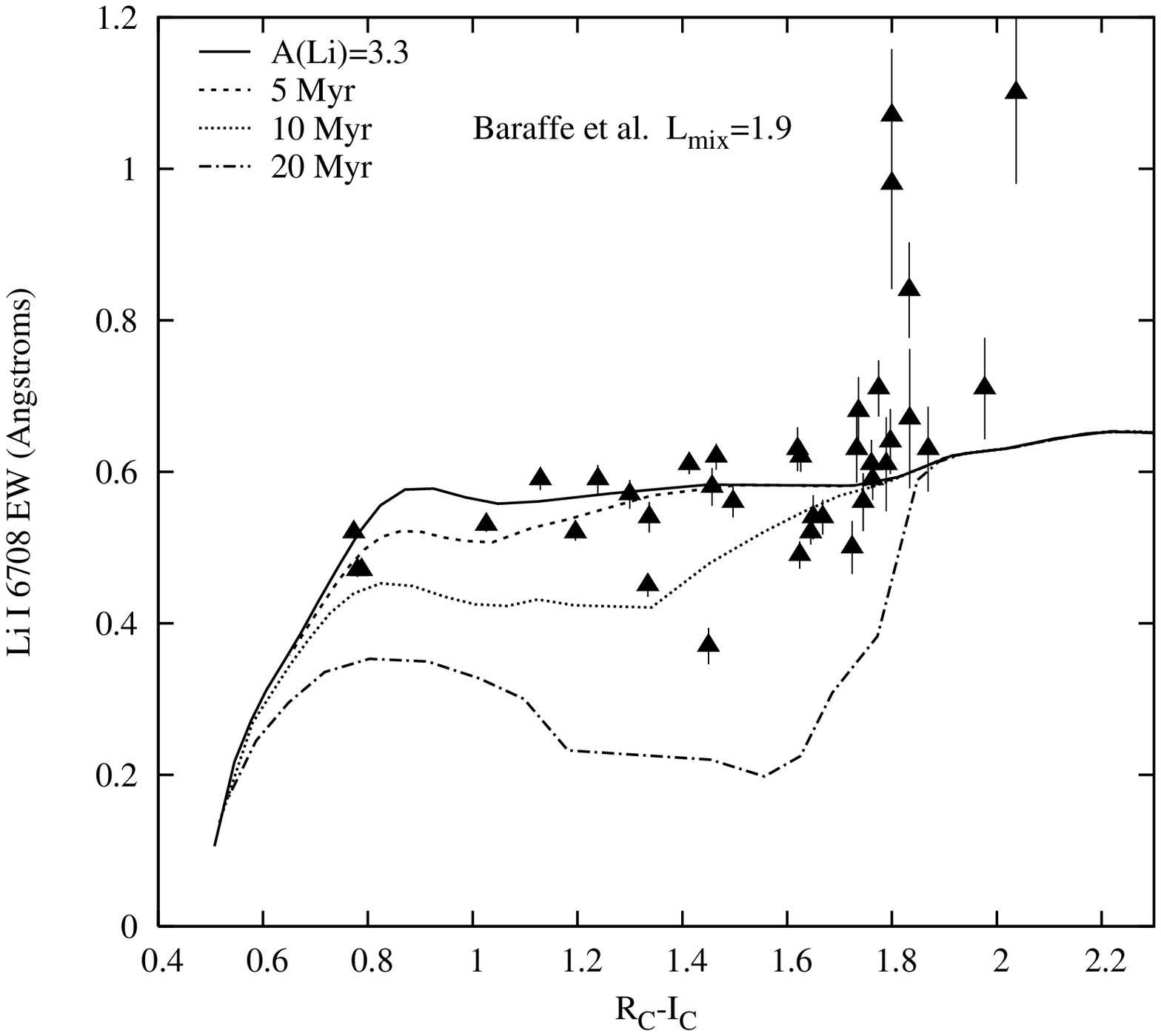}
\end{minipage}
\begin{minipage}[t]{0.45\textwidth}
\includegraphics[width=71mm]{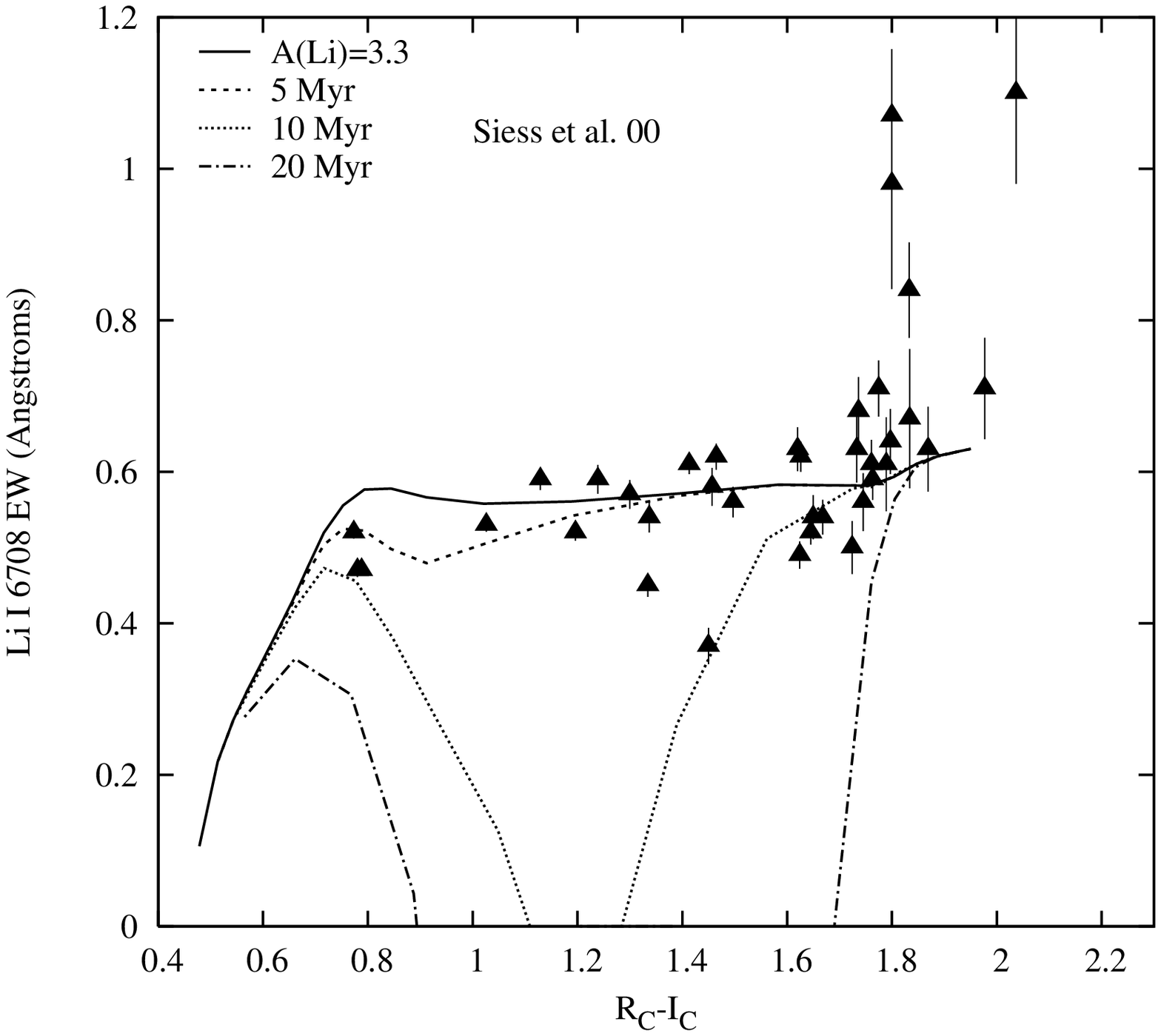}
\end{minipage}
\begin{minipage}[t]{0.45\textwidth}
\includegraphics[width=71mm]{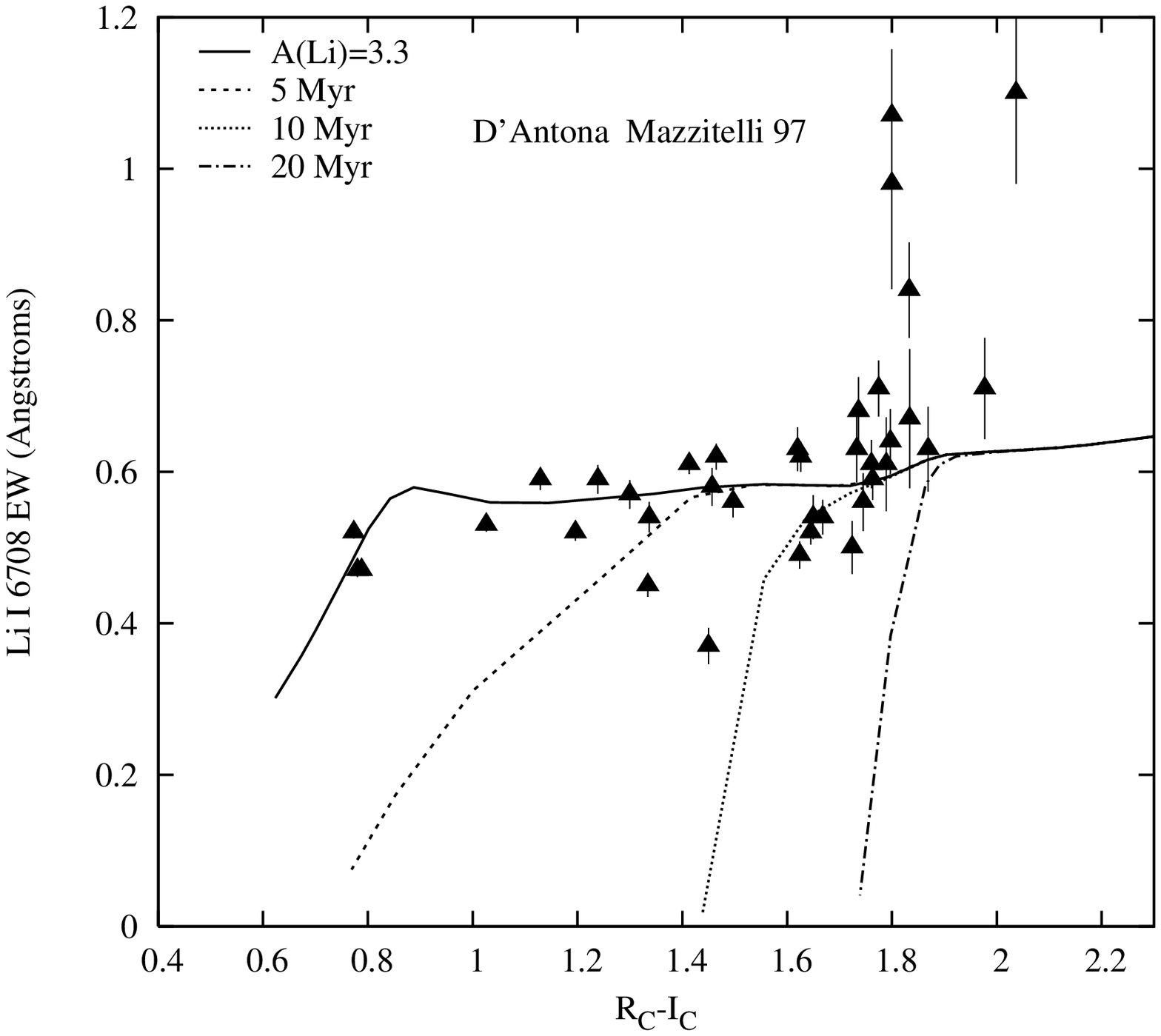}
\end{minipage}
\caption{Isochrones (0, 5, 10, and 20\,Myr) of EW[Li] versus $R_C-I_C$ derived from several
  different models and compared to our observations. An initial Li
  abundance of A(Li)$=3.3$ is assumed and abundances are converted to
  EWs using the relationships defined in Zapatero Osorio et al. (2002)
  and Jeffries et al. (2003) (a) Baraffe et al. (1998, 2002) using a
  mixing length of 1.0 pressure scale heights. (b) Baraffe et
  al. (1998, 2002) using a mixing length of 1.9 pressure scale
  heights. (c) Siess et al. (2000) using a metallicity of $Z=0.02$ and
  no overshooting. (d) D'Antona \& Mazzitelli (1997) using a grey
  atmosphere and the full spectrum turbulence treatment of convection.
}
\label{plotliewvsri}
\end{figure*}

That all of the firm cluster members have EW[Li]$>0.3$\AA\ puts strong
constraints on the age of NGC~2169. For clusters with ages$>30$\,Myr,
there is observational evidence that an ``Li depletion chasm'' opens
up, starting with stars at spectral type $\simeq$M2 and widening with
age to encompass spectral types either side (see Jeffries 2006). The
cool side of this chasm, the so-called Lithium Depletion Boundary, is
sharp and almost model independent. It occurs
when the cores of contracting, fully convective stars reach a
temperature sufficient to burn Li. The
warm side of the chasm exhibits a more gradual roll-off with
temperature (see Fig.~\ref{plotliewvsri}). Here the depletion takes
place at the boundary of a radiative core and a receding convective
envelope. The amount of depletion is sensitively dependent to details
of the convection treatment, interior opacities and chemical composition.

In principle, isochrones of Li depletion can be used as an alternative
way to estimate the age of a cluster. The difficulties in doing so are
converting the data in the observational plane (spectral type/colour
and a spectrum or EW[Li]) into the quantities predicted by models
(i.e. $T_{\rm eff}$ and a Li abundance) or vice versa. A further slight
complication is that models predict Li depletion rather than Li
abundance. The initial Li abundance must be assumed, although a value
of A(Li) ($= 12 + \log {\rm N(Li)/N(H)}$) of $3.3\pm 0.1$ appears to
agree with observational constraints from meteorites and very young
(presumably undepleted) stars (Soderblom et al. 1999).

We chose to perform a comparison of data and models in the
observational plane of EW[Li] versus $R_C-I_C$ colour.
Model temperatures were transformed using the same colour-$T_{\rm eff}$
relations required to make the same models match the Pleiades CMD (see
Section~\ref{lowmassage}).  The predicted abundance was transformed into
EW[Li] using $T_{\rm eff}$ and the relationship between abundance and
EW[Li] derived from cool stellar atmospheres and synthetic spectra by
Zapatero-Osorio et al. (2002) and extended to warmer temperatures and
lower abundances by Jeffries et al. (2003). This latter relationship
was derived to interpret spectra with a spectral resolution of 1.68\AA\
and predicts ``pseudo EWs'' with respect to a local
continuum. In this respect it is ideal for interpreting our measured
EWs.

The comparisons with four models are shown in
Fig.~\ref{plotliewvsri}. These are the isochrones arising from the Li
depletion predicted by: the Baraffe et al. (1998, 2002) evolutionary
tracks, using a mixing length of either 1.0 or 1.9 pressure scale
heights respectively; the Siess et al. (2000) models with no convective
overshoot and a mean metallicity of $Z=0.02$ and; the grey atmosphere
models of D'Antona \& Mazzitelli (1997) featuring the full
spectrum turbulence treatment of convection.

Taking the observations at face value, the majority of targets appear
to possess Li at undepleted levels, with one or two objects showing
some evidence of depletion (amounting at most to about a factor of 10
in target 32). On the other hand, some of the coolest objects appear to
show significantly enhanced Li with respect to the assumed initial
abundance. One of these (target 28) has two mutually consistent
measurements taken on different nights. Looking at the bulk of the
objects, the Li measurements can only give upper limits to the cluster
age. On the redward side ($R_C-I_C>1.5$), the lack of a clear cut
lithium depletion boundary implies a cluster age of $<15$\,Myr for all
the models. For $R_C-I_C<1.5$ there is some model dependence due to the
increasing efficiency of convection as we move from the top-left to
bottom-right of Fig.~\ref{plotliewvsri}. The Baraffe models with mixing
length of 1.0 scale heights suggest an age $<10$\,Myr for most stars
with the possibility of a couple of objects as old as 15\,Myr. Models
with increased convective efficiency suggest progressively younger ages
with the bulk of objects consistent with ages of $\leq 5$\,Myr and
maximum ages of about 12, 10 and 8\,Myr respectively for the Baraffe et
al. model with larger mixing length, the Siess et al. model and the
D'Antona \& Mazzitelli model respectively. 

The ages inferred from the Li depletion are in reasonable agreement
with those inferred from the low-mass isochronal fits using the same
models. We do not find strong evidence for examples of anomalously
rapid Li depletion amongst young stars that might imply a problem with
the evolutionary models as suggested by Song et al. (2002) and White \&
Hillenbrand (2005).

\section{Cluster structure and total mass}

\label{clustermass}

\begin{figure}
\includegraphics[width=80mm]{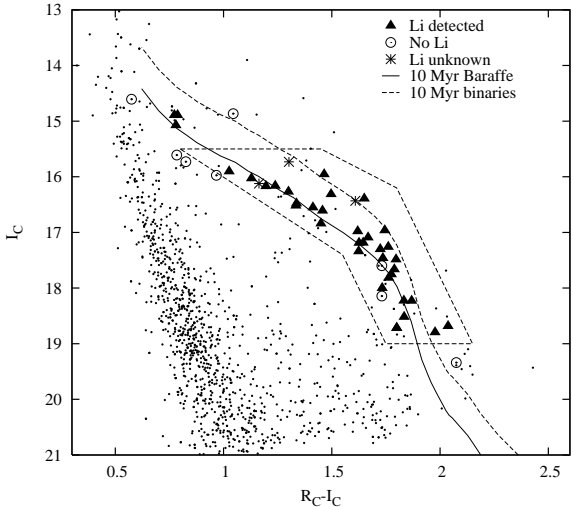}
\caption{A colour magnitude diagram for the area covered by the three
  GMOS fields (see Fig.~\ref{gaiaplot}). Stars with Li (members), those
  without Li (non-members) and those with unknown Li (questionable
  membership) according to Section~\ref{member} are indicated. The
  dashed polygon indicates a region we have chosen for photometric
  selection of cluster members with $15.5<I_{c}<19.0$. Also shown are
  an isochrone and binary sequence from Baraffe et al. (2002 -- with
  mixing length set at 1.0 pressure scale height) corresponding to an
  age of 10\,Myr and intrinsic distance modulus of 10.13 (see Section~\ref{age}).
}
\label{plotiri}
\end{figure}

\begin{figure}
\includegraphics[width=80mm]{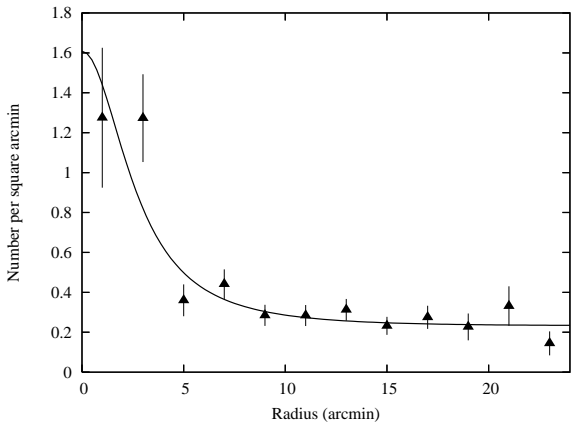}
\caption{Radial surface density profile for photometric cluster
  candidates with $15.5<I_C<19.0$. The best fitting King profile is shown.
}
\label{plotradial}
\end{figure}

Our spectroscopic membership determination can be used to define a
selection region in the $I_C$ versus $R_C-I_C$ CMD that preferentially
includes cluster members and excludes contaminating objects. A larger
sample of probable cluster members can then be chosen from our entire
photometric survey and used to investigate the spatial distribution of
low mass stars in NGC 2169.

Figure~\ref{plotiri} shows a polygon that we have used to select
photometric candidates with $15.5<I_C<19.0$. The appearance of
Fig.~\ref{plotiri} and also Fig.~\ref{cmd1} suggest that few cluster
members could lie redward or blueward of the polygon boundaries. We
have applied this photometric selection criterion to our entire
catalogue (Table~\ref{ccd_catalogue}) and the resulting cluster
candidates are plotted in Fig.~\ref{gaiaplot} to show their spatial
distribution.  There are 64 photometric candidates in the $\sim
70$~arcmin$^{2}$ defined by the overlapping GMOS fields, 38 of which
were among our spectroscopic targets. Of these, 33 are Li-rich cluster
members, 2 are non-members without Li and 3 have uncertain membership
because of the lack of an EW[Li] measurement.  From this we deduce that
the spatial density of contaminating non-members in our selection box
is $\simeq0.05$~stars~arcmin$^{-2}$. Incompleteness due to any lack of
sensitivity in the photometric
survey can be neglected at these magnitudes, although our choice of
photometric selection boundaries, which deliberately exclude
non-members, may have biased the non-member density downwards.

Figure~\ref{gaiaplot} shows a spatial concentration
of photometric cluster candidates. By fitting a Gaussian plus a
constant term to the spatial distribution along the RA and Dec axes we
find the centre of this concentration to be at RA\,$=$ 06h\,08h\,26$(\pm
4)$s, Dec\,$= +$13d\,58m\,23$(\pm18)$s. The radial distribution of the
candidates about this centre is shown in Fig.~\ref{plotradial}. We have
corrected for incompleteness in coverage 
(e.g. due to proximity of bright stars), 
the azimuthal asymmetry in the survey
extent and gaps in the CCD mosaic by normalising with a similar
distribution for stars with $15.5<I_C<19.0$ which are blueward of the candidate
cluster members. The radial distribution has been modelled with a King
profile (see King 1962) plus a constant background density.  The tidal
radius of a cluster in the solar vicinity is given by $r_t =
1.46\,M_c^{1/3}$ (see Pinfield, Hodgkin \& Jameson 1998). We assume
that the mass of the cluster, $M_c = 300\,M_{\odot}$ (see below),
yielding $r_t = 9.8$\,pc, equivalent to 32.2\,arcmin for a distance
modulus of 10.1 (see Section~\ref{age}).  
The best fit core radius, normalisation and
background density are quite insensitive to $r_t$ and hence very
insensitive to the assumed cluster mass.  The best fit King profile
(shown in Fig.~\ref{plotradial}) has a core radius of $(2.8\pm
1.0)$\,arcmin (equivalent to 0.85\,pc at a distance modulus of 10.1),
a normalisation of $(1.7\pm0.6)$ stars~arcmin$^{-2}$, a constant
background of $(0.23\pm0.03)$~stars~arcmin$^{-2}$ and a total of 87
cluster stars integrated out to the tidal radius.  This
background density is far in excess of that deduced from the
spectroscopy above.

If we believe that the cluster spatial distribution {\it is} well
represented by this King profile, then the majority (approximately 220)
of the 302 photometric candidates plotted in Fig.~\ref{gaiaplot} must
be field stars and we would have been fortunate to observe only 2
non-members among 35 GMOS targets with Li measurements, rather than the
9 that would be predicted on the basis of the fitted background
described above. 

We can also check whether the mass function (MF) of
the cluster is close to what has been seen in other clusters and the
field. The limits of $15.5<I_C<19.0$ correspond closely to mass limits of
$1.0> M/M_{\odot}>0.15$ using a 10\,Myr isochrone from Baraffe et
al. (1998, 2002) and a distance modulus of 10.13 (see
Section~\ref{age}). Choice of evolutionary model and variations of age
and distance within the allowed uncertainties discussed in
Section~\ref{age} can change these limits by $\sim 10$ per cent, but do
not change the basic argument set out below.

Using the universal piecewise power law MF advocated by Kroupa (2001),
which matches data from the field and many open clusters, then
87 stars with $0.15<M/M_{\odot}<1.0$ should be found in a
cluster with 3.9 stars with $2.5<M/M_{\odot}<15.0$, corresponding to
$1.2>M_{V}>-4.1$ according to the 10\,Myr Geneva
isochrone used in Section~\ref{himassage}. 
Fig.~\ref{ubvplot} shows there are at least 12-15
such stars in NGC 2169. Therefore either: (i) most of these high mass
stars do not belong to NGC 2169; (ii) the cluster IMF is deficient by a
factor of $\simeq 3$ in low-mass stars; or (iii) the good match of the
King profile to the spatial distribution is coincidental and most of the
``background'' in Fig.~\ref{plotradial} is a population of low-mass
cluster members extending out to the tidal radius and which is {\it much}
more widely dispersed than the high-mass population. A King profile with
the background fixed at the 0.05~stars~arcmin$^{-2}$ implied by
spectroscopy in the GMOS fields is a poor fit (rejected at $>99.99$ per
cent confidence) to the data.

In our view both (i) and (ii) are unlikely, but confirming
that the widely dispersed low-mass cluster candidates are genuine
members would require further spectroscopy.  
For scenario (ii) the total mass of the cluster
(for $0.15<M/M_{\odot}<15.0$) is $\simeq 150\,M_{\odot}$, whereas for
scenario (iii) the total mass is about $300\,M_{\odot}$ if the Kroupa
(2001) mass function is assumed.

\section{Discussion}

\subsection{The age spread in NGC 2169}

The possibility of age {\it spreads} in young clusters and star forming
regions has been vigorously debated. Whether
molecular clouds can sustain star formation for long periods of time
($\sim 10$\,Myr) or whether star formation is a rapid process that is
essentially completed in a free-fall time ($\sim 1$\,Myr) is related to
whether magnetic fields or supersonic turbulence regulate the gravitational
collapse of the clouds (e.g. Mac Low \& Klessen 2004; Mouschovias,
Tassis \& Kunz 2006).

The age distribution of stars can be deduced from their position in H-R
diagrams. Based on several nearby star forming regions including Taurus
and the Orion Nebula cluster (ONC), the claim has been made that star
formation accelerates over the course of $\sim 10$\,Myr in a typical
molecular cloud (Palla \& Stahler 2000, 2002). These claims are
disputed by Hartmann (2001, 2003) who explain apparent age spreads and
accelerating formation rates in the H-R diagram as due to binarity,
variability, dispersion in extinction and accretion properties and
contamination by foreground non-members (see also the discussion of
variability in Burningham et al. 2005).  Palla et al. (2005) have
recently bolstered the idea of a significant age spread with the
observation of several objects in the ONC that may have depleted their
photospheric Li by factors of 3--10. They argue that such depletion
could not occur unless these stars were at least 10\,Myr old and that
such ages are roughly in agreement with their positions in the H-R
diagram. The possible problem we see here is that Palla et al. (2005)
needed to ``unveil'' their spectra prior to determining the Li
abundances, implying that the objects were heavily accreting. It is not
clear that a plane parallel, LTE model will satisfactorily yield Li
abundances in these circumstances.  The \lii~6708\AA\ resonance doublet
forms close to the top of the atmosphere and is vulnerable to NLTE
effects such as overionisation by a non-photospheric UV continuum that
could weaken EW[Li] (e.g. Houdebeine \& Doyle 1996)

At an age of $9\pm 2$\,Myr NGC~2169 is a fascinating cluster with
which to test some of these ideas. First, accretion appears to have
ceased (see Section~\ref{accrete}), but the cluster is young enough
that small changes in age still result in significant changes in
luminosity for low-mass stars.  Hence the scatter of stars in the
H-R diagram will contain information on any age spread
if it can be separated from scatter caused by binarity, intrinsic
variability and differential reddening. Second, Li depletion only begins in
low mass stars after about 5--10\,Myr, so for a given age spread 
any spread in Li depletion should become much more pronounced
in NGC~2169 than in a younger cluster like the ONC.

In Section~\ref{lowmassage} we found that a magnitude-independent
uncertainty of 0.03 mag needed to be added to the uncertainty in each
photometric band in order for the Baraffe et al. (1998, 2002) models to
provide a reasonable fit to the low-mass PMS.  There
could be several contributions to the requirement for this additional
uncertainty: (i) that the isochrone shapes do not represent the data
very well; (ii) variability due to chromospheric activity and starpots;
(iii) differential reddening; (iv) incorrect assumptions about the
binary frequency or mass ratio distribution; (v) an age spread.  Of
these, (i) appears not to be an issue (see Fig.~\ref{bestfits}a), (iii)
is probably limited to less than 0.014 mag scatter in $E(R_{C}-I_{C})$
(Delgado et al. 1992) and (iv) has little effect when changed within
reasonable limits.  Instead it seems that there is additional scatter
about the best fitting isochrone (especially towards the low-mass end)
that could be caused by a combination of (ii) and (v), but also
includes a contribution of 0.01 mag from systematic
photometric uncertainties (see Section~\ref{ccdphotom}).

The additional scatter corresponds to {\it at most} $\pm0.04$ mag
(1-sigma) in $R_C-I_C$. Over the mass range of our cluster members,
isochrones over a small age range are nearly parallel and a 0.04 mag
dispersion is equivalent to only $\pm 1.2$\,Myr when translated into a
shift in age (independent of which models are chosen). As young stars
are known to be variable this must represent an upper limit to how much
of the scatter in the CMD can actually be attributed to an age
spread. Of course one could relax some of the (we believe very
reasonable) assumptions about the binary frequency and mass ratio
distributions to increase this, but even without any binary systems,
the total age spread would be less than 10\,Myr. A further
concern might be that our spectroscopic target selection has prevented
the inclusion of older cluster members that lie below the targeted
cluster members in the CMD. Figure~\ref{plotiri} shows that there is a
significant gap between the cluster members and what are presumably
objects unassociated with the cluster that lie some way below the
cluster main sequence. If they were cluster members, they would need to
be at least 30\,Myr old, which seems an unrealistically large spread.
Our conclusion is that we do not require age spreads beyond a Gaussian
FWHM of 2.5\,Myr to explain the photometric data in NGC 2169.

Supporting evidence for a small age spread comes from the lack of any
large dispersion in the Li abundances. This evidence has the additional
merit of being independent of assumptions about binary frequencies,
differential extinction or variability. Although we have already
expressed our reservations about using the \lii~6708\AA\ line to derive
Li abundances, it is likely that interpretation and modelling problems
could only serve to increase the observed dispersion. Indeed,
Fig.~\ref{plotliewvsri} shows several cooler objects that appear to
have enhanced Li abundances.  The Li observations imply age spreads of
less than 10\,Myr for all the models and as the dispersion must include
a significant component from uncertainties in the Li abundances then
this must be very much an upper limit.  There is just one object
(target 32) that may be significantly Li-depleted and have an age that
is 5--10\,Myr older than all the other cluster members. However, an age
$>5$\,Myr older than the majority of cluster members for target 32 is
not supported by its position in the CMD. Hence we do not find evidence
of Li depletion similar to that found in the ONC by Palla et al. (2005)
that might support an age spread as large as 10\,Myr.

\subsection{Accretion disc lifetimes}

\begin{figure}
\includegraphics[width=75mm]{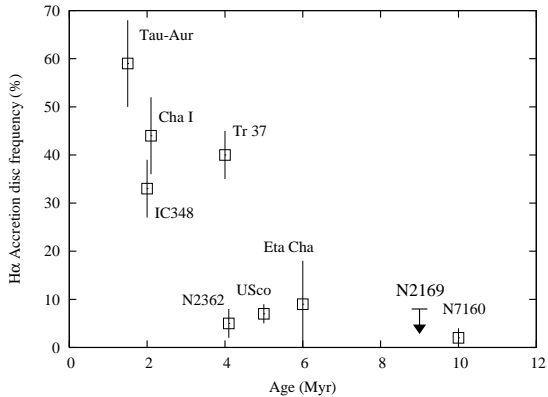}
\caption{The fraction of low-mass stars in NGC 2169 that have accretion
  signatures (according to the White \& Basri [2003] EW[H$\alpha$]
  criteria) compared with other clusters and associations as a function
  of age (derived from the Baraffe et al. [1998, 2002] or Siess et
  al. [2000] models).  The data come from Mohanty et al. (2005) for
  Taurus-Auriga, IC~348, Chamaeleon I and Upper Sco, from Dahm (2005)
  for NGC 2362, from Sicilia-Aguilar et al. (2005) for Tr~37 and
  NGC~7160, and from Jayawardhana et al. (2006) for the $\eta$~Cha
  group. The relative precisions for the cluster ages are typically
  $\pm 1$--2\,Myr.
}
\label{plotaccrete}
\end{figure}

The timescale for the disappearance of accretion signatures in young
low-mass stars is a valuable probe of infalling gas and the evolution
of the inner discs in protoplanetary systems. Broad and strong
H$\alpha$ emission is the most readily available signature
of a strong accretion process. Finding the fraction of stars which
exhibit such a signature as a function of age has been the goal of
several recent investigations (e.g. Mohanty, Jayawardhana \& Basri
2005; Sicilia-Aguilar et al. 2005).  NGC~2169 occupies an important
position, because at a similar age, studies of accretion disc evolution
have largely been confined to sparsely populated nearby moving groups like 
$\eta$~Chamaeleontis and TW~Hya (e.g. Jayawardhana et al. 2006).

All of our NGC~2169 targets have EW[H$\alpha$] below the empirical
accretion thresholds proposed by White \& Basri (2003) and Barrado y
Navascu\'es \& Mart\'in (2003). None of the targets for which we have
the necessary data show a $K$-band near infrared excess indicative of
warm circumstellar material. However, a low-accretion rate or
unfortunate geometric arrangement of the accretion disc and/or flow
could result in a small EW[H$\alpha$] or broadened H$\alpha$ emission
with an EW below these thresholds (e.g. Muzerolle et al. 2000). Only
one object in our sample (target 13) shows a broadened H$\alpha$
emission line that might indicate a low-level of accretion. Comparing
H$\alpha$ EWs with mass accretion rates derived from $U$-band excesses
(Sicilia-Aguilar et al. 2005) suggests that the White \& Basri (2003)
H$\alpha$ EW criterion corresponds to mass accretion rates of $\simeq
10^{-9}\,M_{\odot}$\,year$^{-1}$ for late K and early M stars, although
variations in system geometry, viewing angle and the stellar mass will
blur this boundary.

If we strictly adopt the White \& Basri EW[H$\alpha$] accretion
criterion then the 95 per cent upper limits to the fraction of stars
exhibiting accretion or $K$-band near infrared excess in NGC~2169 are 8
(0/36) and 10 (0/30) per cent respectively. The fraction of accretors
in NGC~2169 is compared to other clusters in Fig.~\ref{plotaccrete}. We
have chosen clusters from Mohanty et al. (2005), Sicilia-Aguilar et
al. (2005), Dahm (2005) and Jayawardhana et al. (2006) where the
fraction of accretors has been (re)determined based on the White \&
Basri (2003) EW[H$\alpha$] criteria, where all the cluster ages have
been determined using the Baraffe et al. (1998) or Siess et al. (2000)
models and where the mass range of the stars considered is similar to
those in NGC~2169.  The data for NGC~2169 strongly reinforce the view
that significant gas accretion ($\ga 10^{-9}\,M_{\odot}$\,year$^{-1}$)
has ceased at ages of 10\,Myr in the vast majority of low-mass stars.

\section{Summary}

The main findings of this paper can be summarised as follows:
\begin{enumerate}
\item We have uncovered the low-mass population of NGC~2169,
  spectroscopically confirming 36 objects with $0.15<M/M_{\odot}<1.3$ 
  as cluster members on the basis of their Li abundances, 
  H$\alpha$ emission and radial velocities. We provide a catalogue of
  these spectroscopic members and a full catalogue of $R_{C}I_{C}$
  photometry covering 880 arcmin$^2$ around the cluster which contains
  several hundred other photometric candidates (see below).

\item The high mass population of the cluster has been used to estimate
  an intrinsic distance modulus of $10.13^{+0.06}_{-0.09}$\,mag. At
  this distance, isochrone fitting with several low-mass evolutionary
  models yields ages from 5 to 11\,Myr. The age from the Baraffe et
  al. (1998, 2002) and Siess et al. (2000) models, 
  which provide the best description of the
  low-mass data, is $9\pm 2$\,Myr. Age constraints from the main
  sequence turn-off and from estimates of Li depletion in the low-mass
  stars are consistent with this age.

\item Using reasonable assumptions for the binary frequency and mass
  ratio distribution, the low-mass isochronal fits do not require any
  age spread in the cluster population beyond a Gaussian FWHM of
  2.5\,Myr. The observed levels of Li depletion are also consistent with
  a small age spread ($<10$\,Myr) and only one M-type cluster member
  shows any evidence of significant Li depletion that might indicate it
  is $>5$\,Myr older than the rest of the cluster. Hence the
  observations do not support scenarios where significant star formation in a
  cluster proceeds over extended periods of time ($\ga 5$\,Myr).

\item On the basis of the strength and width of their H$\alpha$
  emission lines and the lack of any $K$-band near infrared excesses,
  we find no strong evidence of accretion activity or warm circumstellar
  material in the confirmed cluster members. Comparison with
  younger clusters reinforces the idea that significant levels of gas
  accretion cease on timescales $<10$\,Myr for the vast majority
  of low-mass stars.

\item Informed by the spectroscopically confirmed cluster members we
  have photometrically selected several hundred other low-mass cluster
  candidates. A consideration of the number and spatial distribution of
  these candidates suggest either that the cluster has a ``top-heavy''
  mass function or that the cluster's low-mass stars are much 
  more widely distributed than the high-mass stars -- out to radii of
  20 arcminutes. The total cluster mass for stars of
  $0.15<M/M_{\odot}<15$ is 150--300\,$M_{\odot}$.

\end{enumerate}

Although much further away, the low-mass population of NGC~2169 is
larger than those in the kinematically defined groups in the solar
vicinity (e.g. $\eta$~Cha, TW~Hya) which have so far provided the focus
for investigations of the early evolution of stars and planetary
systems at a similar age. Nearby moving groups will continue to provide
the best targets for programmes such as direct imaging, where spatial
resolution is crucial, but clusters like NGC~2169 offer much more
potential for precise statistical investigations of low-mass stellar
properties such as spectral energy distributions, rotation rates and
X-ray activity. The greater distance will frequently be offset (as in
this paper) by the multiplexing capability of multi-object or wide-field
instruments that can observe many low-mass objects simultaneously.

\section*{Acknowledgments}
Based on observations obtained at the Gemini Observatory (program
GN-2005B-Q-30), which is operated by the Association of Universities
for Research in Astronomy, Inc., under a cooperative agreement with the
NSF on behalf of the Gemini partnership: the National Science
Foundation (United States), the Particle Physics and Astronomy Research
Council (United Kingdom), the National Research Council (Canada),
CONICYT (Chile), the Australian Research Council (Australia), CNPq
(Brazil) and CONICET (Argentina).  Also based on observations made with
the Isaac Newton Telescope which is operated on the island of La Palma
by the Isaac Newton Group in the Spanish Observatorio del Roque de los
Muchachos of the Instituto de Astrofisica de Canarias.

This publication makes use of data products from the Two Micron All Sky
Survey, which is a joint project of the University of Massachusetts and
the Infrared Processing and Analysis Center/California Institute of
Technology, funded by the National Aeronautics and Space Administration
and the National Science Foundation. NJM acknowledges the receipt of a
studentship funded by the UK Particle Physics and Astronomy Research
Council.

\nocite{burningham05}
\nocite{landolt92}
\nocite{stetson00}
\nocite{schaller92}
\nocite{lejeune01}
\nocite{rastorguev99}
\nocite{meynet93}
\nocite{houdebeine96}
\nocite{white03}
\nocite{cutri03}
\nocite{jth01}
\nocite{jeffries03}
\nocite{naylor02}
\nocite{naylor98}
\nocite{burningham03}
\nocite{baraffe02}
\nocite{baraffe98}
\nocite{dantona97}
\nocite{leggett96}
\nocite{bessell88}
\nocite{kirkpatrick93}
\nocite{zapatero02}
\nocite{siess00}
\nocite{stauffer98}
\nocite{stauffer99}
\nocite{chiosi92}
\nocite{meynet97}
\nocite{meynet00}
\nocite{barradoldb04}
\nocite{barrado03}
\nocite{muzerolle98}
\nocite{rieke85}
\nocite{piskunov04}
\nocite{taylor86}
\nocite{naylor06}
\nocite{mouschovias06}
\nocite{maclow04}
\nocite{mohanty05}
\nocite{sicilia05}
\nocite{jayawardhana06}
\nocite{song02}
\nocite{white05}
\nocite{dahm05}
\nocite{jeffries05}
\nocite{delgado92}
\nocite{sagar76}
\nocite{abt77}
\nocite{perry78}
\nocite{kroupa01}
\nocite{pinfield98}
\nocite{king62}
\nocite{soderblom99}
\nocite{jeffrieslireview06}
\nocite{dean78}
\nocite{lyra06}
\nocite{muzerolle00}
\nocite{jenniskens94}
\nocite{bailer-jones04}
\nocite{randich01}
\nocite{soderblom93pleiadesli}
\nocite{jones96pleiades}
\nocite{oliveira03}
\nocite{pena94}
\nocite{harris76}
\nocite{jerzykiewicz03}
\nocite{liu89}
\nocite{cuffey56}
\nocite{hoag61}
\nocite{duquennoy91}
\nocite{fischer92}
\nocite{schlesinger72}
\nocite{palla00}
\nocite{palla02}
\nocite{palla05}
\nocite{hartmann01}
\nocite{hartmann03}
\nocite{littlefair04}

\bibliographystyle{mn2e}  
\bibliography{iau_journals,master}


\bsp 

\label{lastpage}

\end{document}